\documentclass[10pt, twocolumn]{IEEEtran}
 \usepackage{amsmath,amssymb}
 \usepackage{subfigure}
 \usepackage{graphicx,graphics,color,psfrag}
 \usepackage{cite,balance}
 \usepackage{caption}
 \allowdisplaybreaks
 \usepackage{algorithm}
 \usepackage{accents}
 \usepackage{amsthm}
 \usepackage{bm}
 \usepackage{algorithmic}
 \usepackage[english]{babel}
 \usepackage{multirow}
 \usepackage{enumerate}
 \usepackage{cases}
 \usepackage{stfloats}
 \usepackage{dsfont}
 \usepackage{color,soul}
 \usepackage{amsfonts}
 \usepackage{cite,graphicx,amsmath,amssymb}
 \usepackage{subfigure}
   \usepackage{booktabs}
 \usepackage{fancyhdr}
 \usepackage{hhline}
 \usepackage{graphicx,graphics}
 \usepackage{array,color}
 \usepackage{multirow}
 \usepackage{graphicx}

 \usepackage{amsmath}

\usepackage{lipsum}
\usepackage{mwe}

\usepackage[dvipsnames]{xcolor}


\newcounter{parentnumber}

\newtheorem{remark}{\bf Remark}
\def\phi{\varphi}

\def\({\left(}
\def\){\right)}

\def\b0{{\mathbf{0}}}

\begin{document}

\title{\Huge Combinatorial Data Augmentation: \\ A Key Enabler to Bridge Geometry- and Data-Driven WiFi Positioning}

\author{Seung Min Yu, Jihong Park, and Seung-Woo Ko\\
\thanks{S. M. Yu is with Korea Railroad Research Institute, Uiwang 16105, South Korea (e-mail: smyu@krri.re.kr). J. Park is with Deakin University, Geelong, VIC 3220, Australia (e-mail: jihong.park@deakin.edu.au). S.-W. Ko is with Inha University, Incheon 22212, Korea (e-mail: swko@inha.ac.kr).  This paper was presented in part at IEEE VTC Spring 2022 \cite{CDA2022_VTC}.}
}

\maketitle

\begin{abstract}
Due to the emergence of various wireless sensing technologies, numerous positioning algorithms have been introduced in the literature, categorized into \emph{geometry-driven positioning} (GP) and \emph{data-driven positioning} (DP). These approaches have respective limitations, e.g., a non-line-of-sight issue for GP and the lack of a labeled dataset for DP, which can be complemented by integrating both methods. To this end, this paper aims to introduce a novel principle called \emph{combinatorial data augmentation} (CDA), a catalyst for the two approaches' seamless integration. Specifically, GP-based datasets augmented from different combinations of positioning entities, called \emph{preliminary estimate locations} (PELs), can be used as DP's inputs. We confirm the CDA's effectiveness from field experiments based on WiFi \emph{round-trip times} (RTTs) and \emph{inertial measurement units} (IMUs) by designing several CDA-based positioning algorithms. First, we show that CDA offers various metrics quantifying each PEL's reliability, thereby filtering out unreliable PELs for WiFi RTT positioning. Second, CDA helps compute the measurement covariance matrix of a Kalman filter for fusing two position estimates derived by WiFi RTT and IMUs. Third, we use the above position estimate as the corresponding PEL's real-time label for fingerprint-based positioning as a representative DP algorithm. It provides accurate and reliable positioning results, says an average positioning error of $1.51$ (m) with a standard deviation of $0.88$~(m). 
\end{abstract}


\section{Introduction}\label{Introduction}

With the widespread use of mobile devices, pinpointing a user's location, called \emph{positioning}, has become an essential ingredient for location-based services, e.g., navigation and location-based gaming \cite{Bensky2016}. Due to the emergence of various wireless sensing technologies, numerous positioning algorithms have been developed, broadly classified into two approaches. One is \emph{geometry-driven positioning} (GP), estimating a user's location from the intersection among different measurements' geometric representations. The other approach is \emph{data-driven positioning} (DP), exploiting a dataset of various location-dependent features to infer a user's location by matching a small number of features extracted from the dataset. This paper aims to bridge the two via \emph{combinatorial data augmentation} (CDA). Specifically, GP-based datasets augmented from different combinations of positioning entities and subsequent outcomes can be used as DP's inputs and labels. We verify the CDA's effectiveness by tackling various issues in WiFi positioning areas, such as the coexistence of \emph{line-of-sight} (LoS) and \emph{non-LoS} (NLoS) propagations, \emph{pedestrian dead reckoning} (PDR), and \emph{fingerprint-based positioning}~(FBP).
\subsection{Prior Work}\label{Introduction:PriorWork}

We review GP and DP approaches by summarizing their properties, representative examples, and limitations.

\subsubsection{GP Approach} This approach utilizes a radio signal's specific physical property captured by the corresponding \emph{positioning element} (PE). For instance, time-based PEs use one fundamental physics theory that a signal constantly propagates with light speed $c\approx 3 \cdot 10^8$ (m/s). In other words, measuring a time-based PE is equivalent to estimating the distance to the user. It can be transformed into the geometry of the user's possible location, e.g., a circle for \emph{time-of-arrival} (ToA) or \emph{round-trip time} (RTT), and a hyperbola for \emph{time-difference-of-arrival} (TDoA).  
In addition to time-based ones, power-based, phase-based, and frequency-based PEs are usable in the same vein, 
summarized in numerous surveys such as \cite{Laoudias2018}~and~\cite{Zafari2019}.

With multiple PEs, we can estimate the user's location by finding the intersection of each PE's geometry, called multilateration. Due to its simple implementation and low operation cost, various multilateration algorithms have been adopted in practical systems. \emph{Observed TDoA} (OTDoA) is a representative positioning algorithm in cellular systems \cite{Fischer2014}. It uses multiple TDoAs of positioning reference signals from different positioning anchors, assuming all anchors are synchronized. An RTT-based localization using WiFi \cite{Han2021} or \emph{ultra-wideband} (UWB) \cite{Feng2020} is another popular method suitable to a scenario without anchor synchronization (e.g., an indoor case) by exchanging positioning request messages and acknowledgments. On the other hand, an inevitable delay occurs, such as $100$-$120$ (msec) for three WiFi RTT measurements \cite{Banin2017}.

A GP approach can provide a decent positioning accuracy when all PEs are measured under LoS conditions, while its accuracy is degraded if a few of them become NLoS  \cite{Ibrahim2018}. An NLoS propagation significantly deviated from a direct path makes the distance to the user overestimated, resulting in inaccurate positioning far from the ground truth. 

Two kinds of GP approaches have been considered to address the NLoS issue. First, there have been attempts to exploit NLoS paths' geometries in GP-based algorithms. In \cite{miao2007positioning} and \cite{han2018sensing}, each NLoS path is characterized by the combination of a time-based PE, \emph{angle-of-arrival} (AoA), and \emph{angle-of-departure} (AoD), assuming all NLoS paths being single-bounce. On the other hand, this assumption is unlikely to be feasible in complicated surroundings with numerous reflectors and blockages like urban areas~\cite{Ko2021_VP_Mag}. Second, PDR exploits mobility information detected by \emph{inertial measurement units} (IMUs) to mitigate the effect of NLoS \cite{Yang2015}. Its core process is to quantify the NLoS effect on positioning results. Conventional approaches rely on a predetermined stochastic distribution (e.g., noise covariance matrices for \emph{Kalman filter} (KF) \cite{Zhang2020}) without concerning current noisy levels.

\subsubsection{DP Approach} Following the recent advancement of \emph{machine learning} (ML) and big data analytics, DP has received significant attention to cope with the NLoS issue mentioned above, thanks to its robustness to measurement perturbations by exploiting many data points and hidden location-dependent features. Specifically, using many PEs as a dataset, ML aims to find the best mapping function between each PE and the corresponding ground-truth label. We can annotate different labels depending on the concerned purpose of using DP. Two representative directions are introduced~below.

First, consider a binary label representing LoS and NLoS propagations. In that case, the resultant DP algorithm is a classifier to identify between the two, helping accurate localization by excluding or mitigating the NLoS effect. In \cite{Marano2010}, a support vector machine is used to identify the LOS/NLoS of UWB by extracting several features from its multi-path profile, including energy and delay spreads. A WiFi signal's LoS/NLoS identification is tackled in \cite{Choi2018}, where its finite bandwidth makes it challenging to resolve multiple signal paths. Instead, a series of \emph{channel state information} (CSI) is used as a feature of a recurrent neural network.  In \cite{Huang2020}, a power angle spectrum containing AoA and AoD is considered a feature of several ML models capable of capturing an NLoS propagation's angular spread.

Second, suppose we label a user's coordinates on which the corresponding PEs are collected. Then, the resultant DP algorithm becomes a regression to estimate the user's location directly, called FBP \cite{Vo2015}. A different PE has been used as a fingerprint, such as received signal strength \cite{Yiu2017}, CSI \cite{Wang2016TVT}, and RTT \cite{Hashem2020}. However, a single fingerprint approach is prone to the wireless environment's slight changes due to weather, user density, and mobility. Thus, it is recommended to use heterogeneous fingerprints \cite{Zhou2021, Li2019} or cooperate with nearby devices \cite{Chen2016}, making the algorithm more robust against dynamic environment change.

It is noteworthy that most DP algorithms mentioned above are built on the principle of supervised learning, requiring the label of every training data sample. As exemplified in annotating an LoS or NLoS propagation for their identification and the location's coordinates for every fingerprint, data labeling is a time-consuming and labor-intensive campaign. Crowdsourcing is a new idea to relieve such burdens by allowing ordinary users to participate in labeling their measurements \cite{Wang2016}. On the other hand, the crowdsourced data label accompanies an unpredictable error, limiting the DP algorithm's accuracy and reliability due to noisy data labels.

A few works in the literature incorporate geometry information into DP to generate labeled data samples. In \cite{Caso2015}, a multi-wall multi-floor propagation model is calibrated using prior knowledge of the indoor environment to generate many virtual fingerprints at unmeasured locations. In \cite{Renaudin2018}, a ray-tracing simulation is used to generate data samples by constructing the indoor environment of the concerned area according to the given floor map. In \cite{Silva2022}, \emph{support vector regression} (SVR)-based data augmentation algorithm is designed by exploiting the obstacle information at the concerned location. On the other hand, the above works' operation relies on the concerned environment's prior information, e.g., floor map and obstacle locations, which are difficult to be obtained, especially when arriving at a new site.

\subsection{Contributions}

GP and DP have their respective advantages and disadvantages. For instance, as opposed to DP, which requires many training samples, GP is instantly applicable without training. Notwithstanding, GP is sensitive to LoS conditions, in contrast with DP, which is robust even under NLoS conditions. Given their complementary advantages, it is natural to combine both approaches, which are yet non-trivial for the following reasons. First, GP can operate with a relatively small number of PEs; for example, at least three RTTs are required for a unique positioning, while DP requires massive data samples to avoid overfitting. Second,  GP can derive the location estimate from PEs without the ground-truth label. In contrast, DP relies on labeled data samples, making it challenging to use PEs directly in a supervised learning-based DP algorithm

To address the above issue, we propose a novel concept of CDA, which is a bridge for seamless integration between the two. Specifically, CDA utilizes GP to create a large volume of data samples from different PE combinations, each of which embeds useful information to infer the user's location. Then, it is ready to use data-centric algorithms for precise localization, such as feature extraction and data filtering. Besides, we can tackle DP's labeling issue by annotating the augmented data samples with the resultant location estimates, which are reliable enough to design a practical DP with acceptable accuracy. Such CDA's effectiveness is extensively verified by field experiments using WiFi RTTs and IMUs. To our knowledge, it is the first work attempting to combine the two. The main contributions of this work are summarized~below.

\begin{itemize}
\item \textbf{CDA-Based Data Filtering}: Among the augmented data samples, we can filter out unreliable ones severely affected by the NLoS effect. Two byproducts obtained in the CDA process are used to this end: residual error and RTT sum. We design a tandem filtering algorithm based on the two metrics. The remaining data samples after the filtering are used for positioning, reducing the average positioning error from $6.69$ (m) to $1.77$~(m).    
\item \textbf{CDA-Based Data Fusion}: CDA helps evaluate the above location estimate's accuracy without the ground truth since the augmented data samples' spatial variance is highly correlated to the accuracy. The resultant evaluation is helpful when fusing multiple location estimates obtained from different PEs, i.e., {integration of PE-based and IMU-based positioning via KF.} It is experimentally shown that the resultant positioning error's mean and standard deviations are $1.65$ (m) and $1.01$ (m), whereas those using the conventional method are $1.69$ (m) and $1.21$ (m). 
\item \textbf{CDA-Based Labeling}: The location estimates using the above method are accurate enough to be used as labels for DP. The idea is verified by FBP, whose positioning error's mean and standard deviation can reach $1.51$ (m) and $0.88$ (m), respectively, which outperform labels' performance [i.e., mean $1.65$ (m) and standard deviation $1.01$ (m)]. 
\end{itemize}

The remainder of the paper is organized as follows. Sec.~\ref{Sec2} introduces the concept of CDA with its features for overcoming the drawbacks of standalone GP and DP. Sec.~\ref{Sec3} and \ref{sec4} present 
several positioning techniques derived from CDA, including reliability-based data filtering, real-time measurement covariance matrix update, and real-time labeling for FBP. Lastly, we conclude the work in Sec.~\ref{Sec6}.

\section{A Primer on Combinatorial Data Augmentation}\label{Sec2}

\begin{figure*}
\centering 
{\includegraphics[width=15cm]{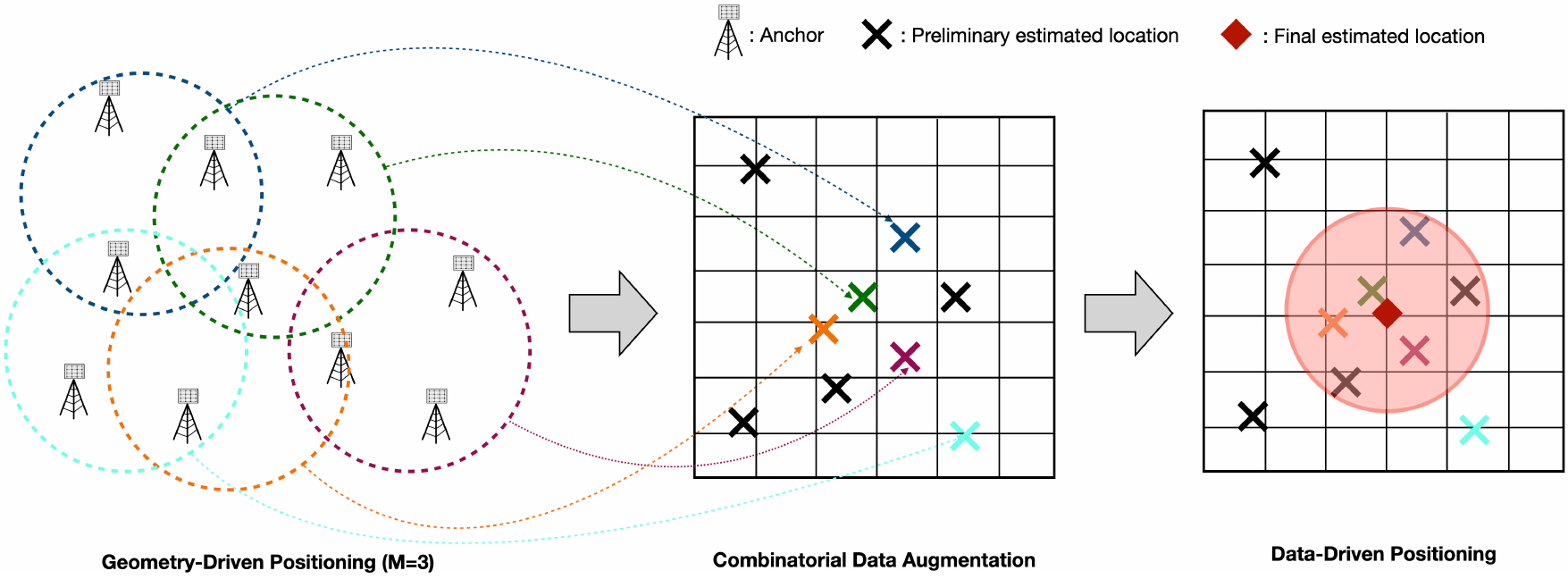}} 
\caption{Graphical representation of an integrated positioning approach using CDA. }\label{Overview_CDA}
\end{figure*}

\begin{table}[]\caption{Notations and Definitions}\label{Table:Notation}
\centering
\begin{tabular}{ll}
\toprule
Notations                                   & Definitions                              \\ \midrule
$N$                                         & Number of APs                           \\
$\boldsymbol{p}_n$                          & Location of AP $n$                      \\
$\boldsymbol{x}(t)$                         & Mobile's location at time $t$           \\
$\boldsymbol{x}[k]$                         & Location of MP $k$                      \\
$\boldsymbol{\tau}[k]$                      & RTTs at MP $k$                          \\
$L$                                         & Number of APs for deriving one PEL      \\
$\boldsymbol{z}_{\ell}$                     & PEL $\ell$                              \\
$\mathcal{Z}$                               & Set of PELs                             \\
$\alpha(\ell)$                              & RE of PEL $\ell$                        \\
$c$                                         & Light speed                             \\
$\beta(\ell)$                               & RS of PEL $\ell$                        \\
$q$                                         & The portion of remaining PELs           \\
$\boldsymbol{y}_{\textrm{RE}\&\textrm{RS}}$ & Position estimate of RE \& RS filtering \\
$\boldsymbol{y}_{\textrm{KF}}$              & Position estimate of CDA-based KF\\
$\mathbf{G}$                &Kalman gain\\ 
$\mathbf{R}$                &Measurement covariance matrix\\ 
$\mathbf{Q}$                &Prediction covariance matrix\\ 
$\boldsymbol{s}^{(e)}_i$  &Data sample $e$'s $i$-th input feature\\
$\boldsymbol{y}^{(e)}_j$  &Data sample $e$'s $j$-th real-time label\\
\bottomrule      
\end{tabular}
\end{table}

\subsection{WiFi Positioning and Problem Formulations}\label{Subsec:SystemModel}

Consider a wireless network comprising multiple WiFi \emph{access points} (APs) and a user holding his smartphone. The sets of WiFi APs are denoted by $\mathcal{N}=\{1,\cdots, N\}$. All APs are stationary, and their locations are assumed to be known without loss of generality. Each AP's \emph{two-dimensional} (2D) coordinates are denoted by $\boldsymbol{p}_n\in \mathbb{R}^{2\times 1}$, $n\in\mathcal{N}$. On the other hand, the user is mobile. 
His 2D coordinates at time $t$, defined as $\boldsymbol{x}(t)\in \mathbb{R}^{2\times 1}$, are unknown with no prior knowledge like relevant statistics and surrounding information.  

We use RTT as a primary PE. The user's smartphone can measure RTTs from multiple APs using a \emph{fine timing measurement protocol} (FTM), verified to provide precise RTT estimates at picosecond granularity in LoS conditions~\cite{Ibrahim2018}. Denote $\{t_k\}$ a sequence of time instants when RTTs are measured. The location where the RTTs are measured at $t_k$ is defined as a \emph{measurement point} (MP)~$k$, whose coordinates are defined as  $\boldsymbol{x}[k]=\boldsymbol{x}(t_k)$. The set of MPs is denoted by $\mathcal{K}=\{1,\cdots, K\}$. 
Each user collects RTTs at MP $k$, denoted by $\boldsymbol{\tau}[k]=\left[\tau_1[k], \cdots, \tau_{N}[k]\right]^T\in\mathbb{R}^{N \times 1}$, leading to the following two positioning problems:  
\begin{enumerate}
\item Given $\boldsymbol{\tau}[k]$, the user attempts to estimate his location at MP $k$, defined as ${\boldsymbol{y}}[k]=f_1\left(\boldsymbol{\tau}[k]\right)$, where 
\begin{align}\label{GP_Def}\tag{P1}
f_1=\arg\min_{f:\mathbb{R}^{N}\rightarrow \mathbb{R}^2} \left\|{f(\boldsymbol{\tau}}[k])-{\boldsymbol{x}}[k]\right\|_1.  
\end{align}
Here, $\|\boldsymbol{x}\|_p$ represents a $p$-norm of $\boldsymbol{x}$. 
\item Given $\{(\boldsymbol{\tau}[k], \boldsymbol{y}[k])\}$, we can train a supervised learning-based ML model $f_2$ for FBP, mapping the user's RTTs onto the corresponding location estimates in \ref{GP_Def}, given~as 
\begin{align}\label{fingerprint_Def}\tag{P2}
f_2=\arg \min_{f:\mathbb{R}^{N}\rightarrow \mathbb{R}^2} \sum_{k\in\mathcal{K}} \mathsf{Loss}\left({\boldsymbol{y}}[k], f(\boldsymbol{\tau}[k])\right),
\end{align}
where $\mathsf{Loss}$ is the loss function depending on the concerned ML model. 
\end{enumerate}
The above two problems are cascaded such that we input each user's location estimates in \ref{GP_Def}, say ${\boldsymbol{y}}[k]$, into \ref{fingerprint_Def}. If $\boldsymbol{y}[k]=\boldsymbol{x}[k]$ for all $k\in\mathcal{K}$, \ref{fingerprint_Def} is equivalent to a conventional FBP with the ground-truth label. Therefore, the paper's central theme is to solve \ref{GP_Def} by coping with the issue of NLoS, whose effectiveness is verified by \ref{fingerprint_Def}. 
Noting that \ref{GP_Def} is a snapshot problem at a specific MP, we omit the index $k$ for brevity unless specified.  All notations in this work are summarized in Table \ref{Table:Notation}.

\subsection{A Principle of Combinatorial Data Augmentation}
The goal of this subsection is to introduce CDA, a key enabler for solving \ref{GP_Def}. 
For ease of notation, we introduce a function $g_{{M}}:\mathbb{R}^{M}\rightarrow \mathbb{R}^2$, a conventional GP algorithm returning a location estimate by inputting $M$ APs' RTTs. Specifically, denote $\mathcal{M}$ a set including the concerned $M$ APs. Then, the function $g_{{M}}$ finds the outcome $\boldsymbol{z}=g_{M}(\{\tau_m\}_{m\in\mathcal{M}})\triangleq g_{M}(\boldsymbol{\tau}, \mathcal{M})\in \mathbb{R}^{2 \times 1}$, the resultant location estimate minimizing a given criterion. For example,  \emph{linear least square through reference selection} (LLS-RS) is a representative GP algorithm targeting to follow the well-known \emph{least square} (LS) structure \cite{Guvenc2008}. 
In an ideal case with no NLoS path and no measurement error,  
the resultant LS error is always zero if at least three RTTs are given ($M\geq 3$). In reality, on the other hand, various obstacles and blockages render the error significant. 

Next, we introduce CDA with Fig. \ref{Overview_CDA} illustrating its example, exploiting many AP combinations to augment a real-time dataset. Specifically, all  possible combinations of $M$ APs among the entire $N$ ones are grouped as $\{\mathcal{M}_{\ell}\}_{\ell=1}^L$ with $\mathcal{M}_{\ell}$ being the $\ell$-th combination and $L={N \choose M}$. Each AP selection is one-to-one mapped to the location estimate via the function $g_M$, given as
\begin{align}\label{Each_PEL}
\boldsymbol{z}_{\ell}=g_{{M}}(\boldsymbol{\tau}, \mathcal{M}_{\ell}), \quad \ell=1,\cdots, L.   
\end{align}
We call $\boldsymbol{z}_{\ell}$ a \emph{preliminary estimated location} (PEL). As shown in the middle subfigure in Fig. \ref{Overview_CDA}, PELs are dispersed on the floor plan since the degree of NLoS effect on each PEL is different. Figuring out such a degree helps find the user's location more precisely, dealt with in the sequel.

\subsection{{Advantages of Combinatorial Data Augmentation}}

The proposed CDA-based positioning approach has the following advantages for overcoming GP and DP's drawbacks mentioned above. 
\subsubsection{{Harnessing the Power of Big Data}} Noting that the number of AP combinations ${N \choose M}$ asymptotically scales as the order of $N^M$, it is reasonable to consider the augmented data as \emph{big data} when the number of APs $N$ is sufficiently large. Consequently, extracting various latent information embedded in the augmented data is possible using several data-analytic techniques, such as clustering, data embedding, and data mining. Some discovered information can guide us to find a user's location more accurately.     

\subsubsection{{Practical Design}} 
CDA utilizes low-complexity GP algorithms for augmenting a sufficient number of data samples based on one feasible prerequisite that APs' positions are given in advance. It is verified by the following field experiments that CDA is effectively implementable on a hand-held device, e.g., a smartphone, whose computation capability is limited.   

\subsubsection{{Compatibility with Existing Algorithms}} While we explain CDA based on RTT-based positioning, 
CDA can work well with other GP algorithms according to available PEs, extending its usage into various applications. For example, when the multi-path profile of a wireless propagation is given in terms of AoA, AoD, and delay as in \cite{han2018sensing}, it is possible to find multiple PELs by selecting a few paths among the entire ones. Besides, CDA helps quantify the current position estimate's reliability from the spatial distribution of PELs, helping PDR to combine multiple position estimates. 
The latter is verified in the sequel.

It is worth noting that a few existing works use similar approaches to CDA. For example,  
a \emph{least median square algorithm} (LMeS) is developed in \cite{Qiao2014} that a PEL with the minimum median square error is considered the user's location. In \cite{Chen1999}, a \emph{residual weighting algorithm} (RWGH) is proposed by more weighting a PEL with a smaller residual error. Despite the similarity, the above algorithms are categorized into GP since they do not reflect each PEL's distinct feature, which is DP's domain. In the following section, we suggest new approaches to unleashing the full potential of PELs and verify the proposed algorithms' superiorities by comparing them with the above prior works.

\begin{figure*}[t]
\centering
\subfigure{
\includegraphics[width=5.5cm]{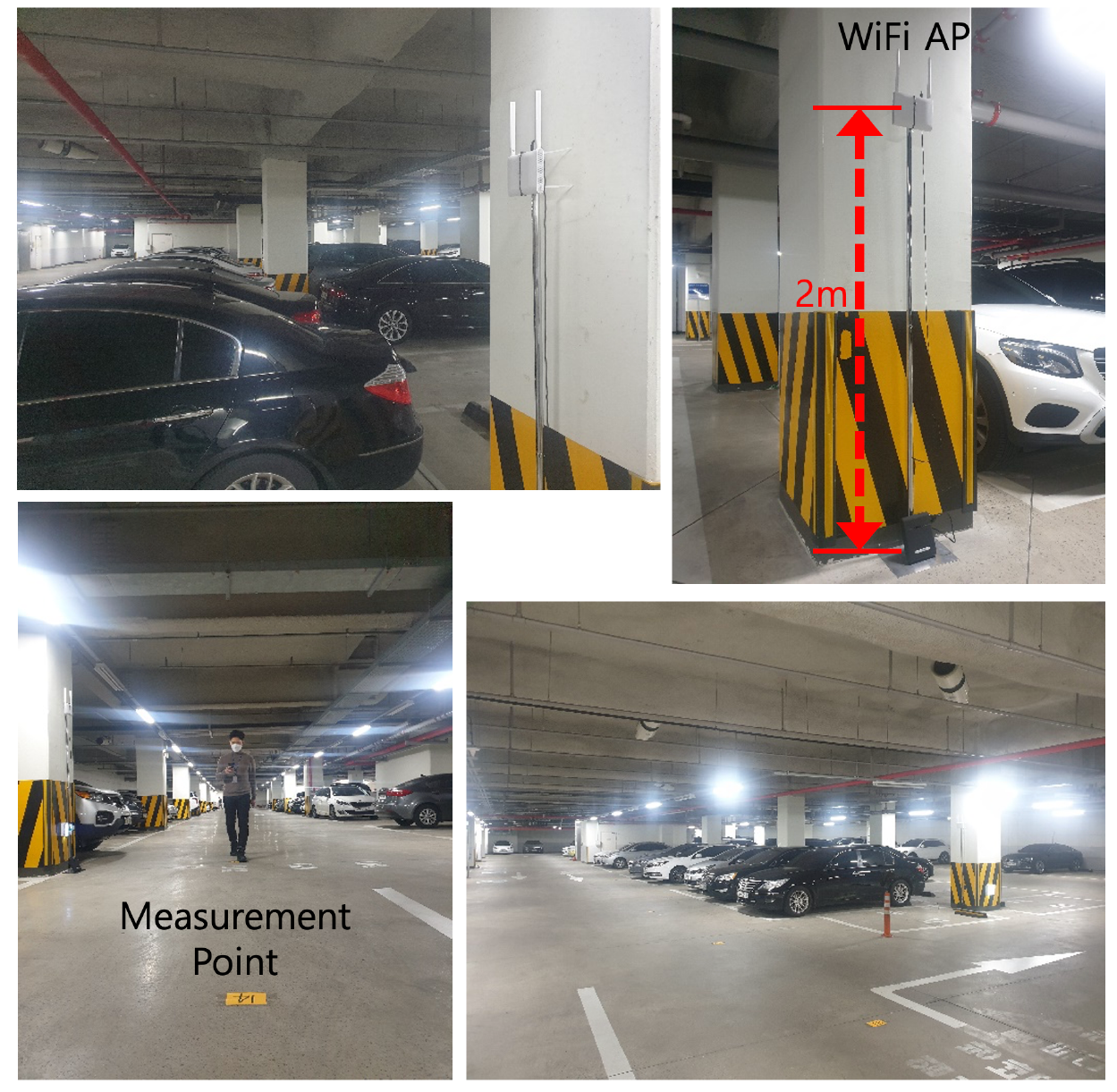}
}
\subfigure{
\includegraphics[width=11.5cm]{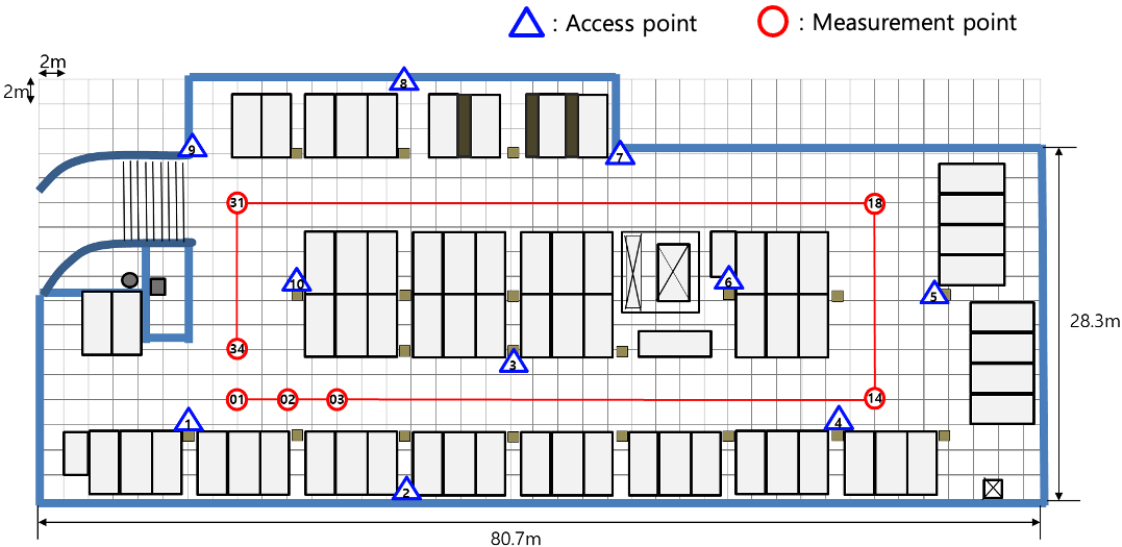}
}
\caption{Experiment photos and the experimental site's floor plan. The detailed experiment settings are described in Sec. \ref{Subsection:Setting}. } \label{ExperimentSite}
\end{figure*}

\section{Combinatorial Data Augmentation Helps WiFi RTT Positioning}\label{Sec3}

This section will showcase how CDA with many PELs helps achieve a precise position estimate for WiFi RTT positioning. 

\subsection{Initial Setting}\label{Subsection:Setting}
We conducted field experiments at the underground parking lot of Building 11 at Korea Railroad Research Institute, Uiwang, Korea.  We use $10$ WiFi APs designed based on Qualcomm IPQ $4018$ ($N=10$) and one Google Pixel2 XL smartphone, both of which support FTM.  We deploy the APs at $2$ meters in height at different locations.  On the other hand, the user holds the smartphone at a height of $1.1$ (m). The user walks around the experiment site, defined as one experiment. 
Each experiment comprises $34$ MPs. At each MP, the smartphone records RTTs and the measurements of built-in IMUs like accelerometers and gyroscopes. We repeated this experiment $12$ times for one hour. {Several experiment photos and the experimental site's floor plan are given in Fig. \ref{ExperimentSite}.}  

For CDA, the number of APs needed for deriving one PEL is set as three ($M=3$). It is the minimum number for unique positioning and provides the best positioning accuracy among all possible numbers, explained in the sequel. Thus, the number of AP selections $L$ specified in \eqref{Each_PEL} becomes ${10 \choose 3}=120$. 
Given each AP selection $\mathcal{M}_{\ell}$, $\ell=1,\cdots 120$, we use the LLS-RS method in \cite{Guvenc2008} as a function $g_{M}$ to derive the corresponding PEL, say $\boldsymbol{z}_{\ell}=g_{3}(\boldsymbol{\tau}, \mathcal{M}_{\ell})$. 
At each MP, we compute a set of PELs as
\begin{align}\label{PEL_set}
\mathcal{Z}=\{\boldsymbol{z}_{1}, \cdots, \boldsymbol{z}_{120}\}.
\end{align}

\begin{figure*} 
\centering 
{\includegraphics[width=18cm]{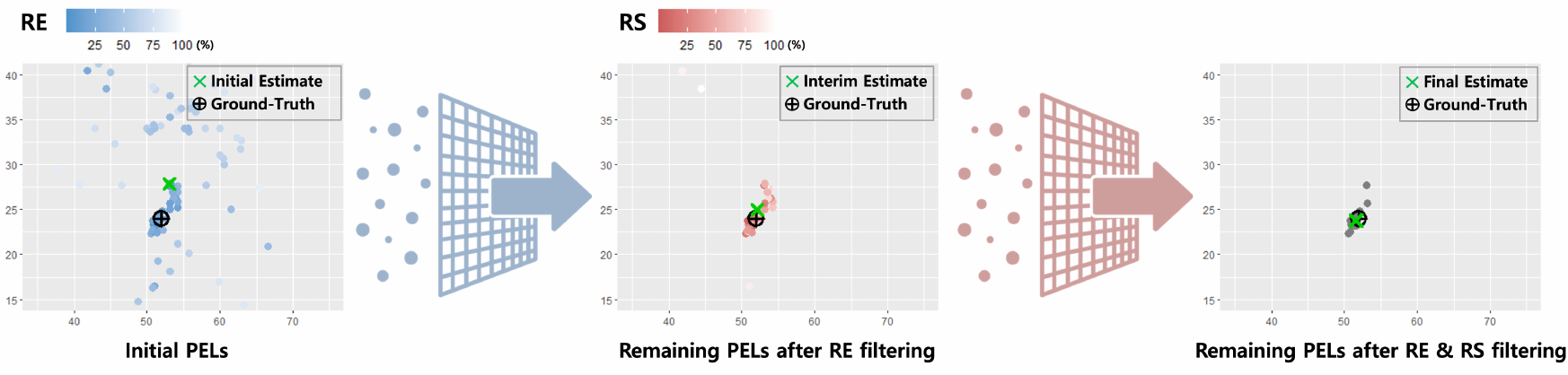}} 
\caption{Graphical illustration of reliability-based PEL filtering at MP $22$, comprising two cascaded filters based on RE and RS values. The number of initial PELs is $120$. The first filter only passes $38$ PELs whose RE values are smaller than the others.  The second filter only passes $12$ PELs among the incoming ones whose RS values are smaller than the others. 
}\label{WiFiPos_wo_PDR}
\end{figure*}

\subsection{Reliability-Based PEL Filtering}\label{subsection:PEL_Filter}
\subsubsection{Algorithm Description}
In this subsection, we attempt to estimate the user's position when $\mathcal{Z}$ is given.
One straightforward way is that their representative value, e.g., median, is considered the user's location estimate. In general, a median-based estimator is known to provide solid performance in many applications since a few outliers highly different from the others are easily ignored \cite{Qiao2014}.  On the other hand, the median of the entire PELs can be significantly far from the ground truth (see the left figure in Fig. \ref{WiFiPos_wo_PDR}). The reason is that many PELs can be severely biased in a particular direction due to strong NLoS propagations made by walls, pillars, and parked vehicles. 

We can overcome the above limitation by picking a few \emph{reliable} PELs less affected by NLoS environments. To this end, we use the following two metrics to quantify the NLoS~effect.  
\begin{itemize}
\item \emph{Residual Error}: Due to measurement errors and NLoS propagations, every PEL $\{\boldsymbol{z}_{\ell}\}$ induced by a different RTT combination is unlikely to meet a single point. In other words, the resultant distances from PEL $\ell$ to APs in $\mathcal{M}_{\ell}$, say $\|\boldsymbol{z}_{\ell}-\boldsymbol{p}_n\|$ with AP $n$'s coordinates $\boldsymbol{p}_n$ specified in Sec. \ref{Subsec:SystemModel}, cannot be the same as those derived from the corresponding RTTs, say $\frac{c\cdot {\tau}_n }{2}$ with $c$ being the light speed. The sum of these errors is called a \emph{residual error}~(RE), given as 
\begin{align}\label{RE_Def}
\alpha(\ell)=\sum_{n\in\mathcal{M}_{\ell}}\left\|\left\|\boldsymbol{z}_{\ell}-\boldsymbol{p}_n\right\|_2-\frac{c\cdot {\tau}_n }{2}\right\|_1.   
\end{align}
A RE $\alpha(\ell)$ can be significant when the RTTs used for deriving the PEL are severely corrupted. As a result, comparing REs helps speculate which PEL is more reliable to represent the ground truth. On the other hand, one occasionally observes that a few PELs with small REs can be placed far from the ground truth. It is thus required to use another metric together. 
\item \emph{RTT sum}: A smaller RTT implies that the user is likelier to be located in an LoS sight from the corresponding AP. It inspires us to establish one hypothesis that a PEL with a smaller \emph{RTT sum} (RS), defined as   
\begin{align}\label{RS_Def}
\beta(\ell)=\sum_{n\in\mathcal{M}_{\ell}}\tau_n, \end{align}
represents a more accurate estimate of the user's location. The hypothesis is well-verified in Appendix A.
\end{itemize}

Using $\{\alpha_{\ell}, \beta_{\ell}\}$, we attempt to filter out unreliable PELs by designing a tandem filter as shown in Fig.~\ref{WiFiPos_wo_PDR}. The detailed procedure is given below. 
\begin{enumerate}
\item \emph{RE Filtering}: First, arrange all PELs in ascending order in terms of RE $\alpha$ \eqref{RE_Def}, e.g., $\alpha(\ell_1)\leq \alpha(\ell_2)$ if and only if  $\ell_1<\ell_2$. Then, the first filter passes only the top $38$ PELs, while the others are discarded (see the middle subfigure in Fig. \ref{WiFiPos_wo_PDR}).  The set of remaining PELs after RE filtering is denoted by $\mathcal{Z}_{\textrm{RE}}$.
\item \emph{RS Filtering}: Second, arrange $\mathcal{Z}_{\textrm{RE}}$ in an ascending order in terms of RS $\beta$ \eqref{RS_Def}, e.g., $\beta(\ell_1)\leq \beta(\ell_2)$ if and only if  $\ell_1<\ell_2$.  Then, the second filters pass only the top $12$ PELs (see the right subfigure in Fig. \ref{WiFiPos_wo_PDR}).  The  set of remaining PELs after RE \& RS filtering is denoted by 
 $\mathcal{Z}_{\textrm{RE}\&\textrm{RS}}$.
\item \emph{Positioning}: Last, the corresponding MP's coordinates are estimated by computing the remaining PELs' median, given as
\begin{align}\label{LocEst_Filtering}
\boldsymbol{y}_{\textrm{RE}\&\textrm{RS}}=\mathsf{median}(\mathcal{Z}_{\textrm{RE}\&\textrm{RS}}).
\end{align} 
\end{enumerate}

Here, we set the portion of the final PELs used in \eqref{LocEst_Filtering} as $q=\frac{|\mathcal{Z}_{\textrm{RE}\&\textrm{RS}}|}{L}=\frac{12}{120}=0.1$, which is jointly configured with another key parameter $M$, explained in the following remark.

\begin{remark}[Parameter Configuration]\label{Remark:ParamConfig} \emph{The parameters $q$ and $M$ should be jointly configured to generate more reliable PELs. To explain, we define a \emph{clean} RTT as one whose error is less than $1$ (m).  
We experimentally investigate that the probability of a clean RTT, denoted by $\delta$, is between $0.4$ and $0.5$ in various indoor environments (e.g., $\delta=0.465$ in the concerned experiment site). 
A PEL is considered reliable if all RTTs used to derive the PEL are clean.
Given $\delta$, the likelihood of a reliable PEL is $\delta^M$, confirming the optimality of the above $q$ setting with $M=3$ (i.e.,  $0.465^3\approx 0.1$). Besides, the expected number of reliable PELs is ${N \choose M} \delta^M$, which is 
maximized when $M=3$ in the current number of APs $N=10$. 
The optimality of the current configuration is experimentally verified in Appendix~B. 
}
\end{remark}

\begin{remark}[Filtering order]\emph{The filtering order between RE and RS is determined based on the proposition that a PEL's reliability depends on the concerned APs' arrangement as well as their LoS conditions. Consider PEL $\ell$ whose all APs in $\mathcal{M}_{\ell}$ are under LoS conditions. It is shown in \cite{Olone2018} that PEL $\ell$'s \emph{Cramer-Rao lower bound} (CRLB) can be affected by the concerned AP's geometry and decreased when APs in $\mathcal{M}_{\ell}$ are distributed across the entire space rather than biased to one direction. Then, PEL $\ell$ is likely nearer to the ground truth, causing a smaller RE $\alpha(\ell)$. On the other hand, the counterpart RS $\beta(\ell)$ of \eqref{RS_Def} could be significant and easily filtered out if we use the RS filter first.  As a result, the policy of RE-first \& RS-following is mainly used to preserve PELs satisfying both.}
\end{remark}

\begin{figure}
\centering 
{\includegraphics[width=9.0cm]{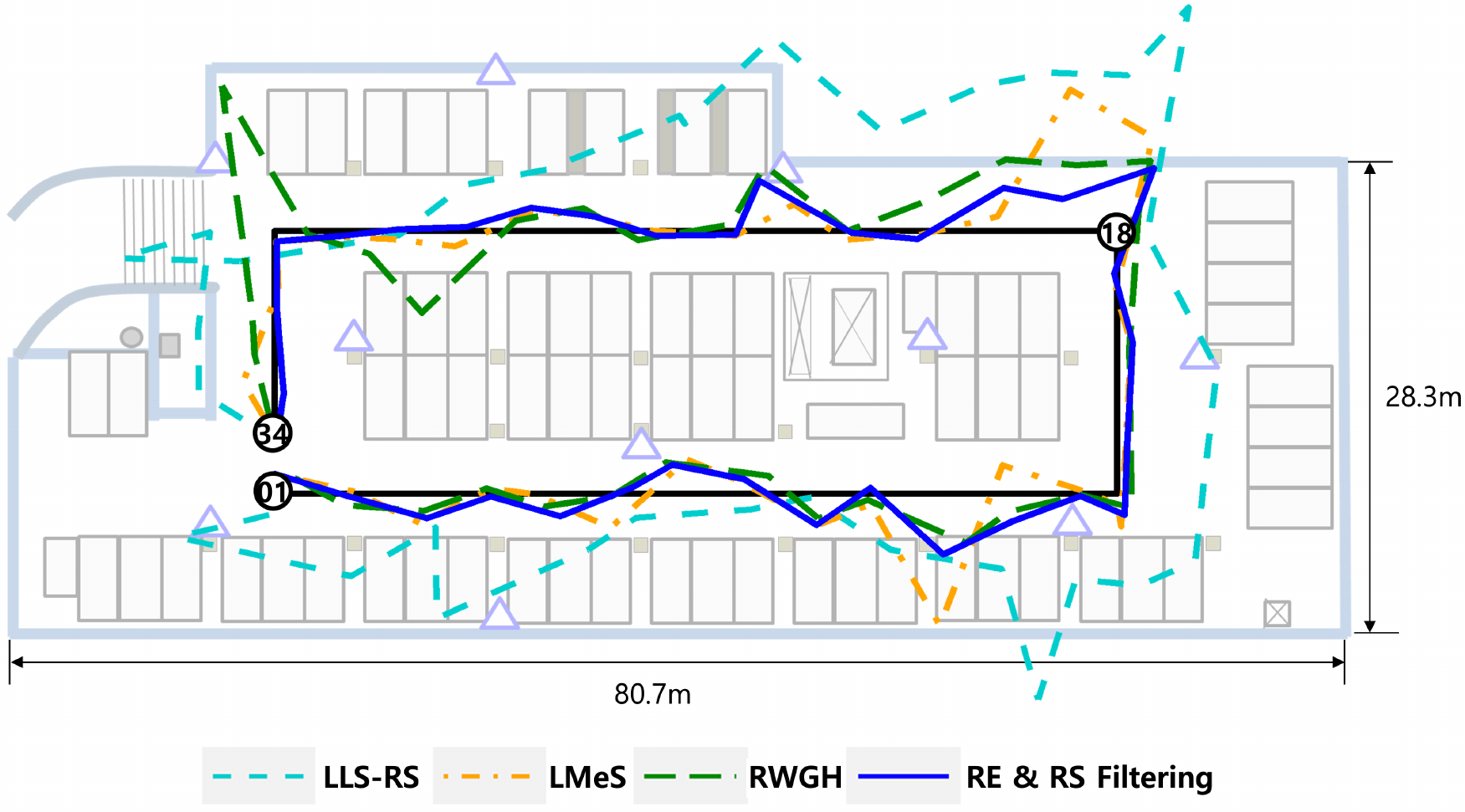}} 
\caption{Graphical representation of location traces when different algorithms are used without PDR, drawn on the floor plan of the experimental site. All results are illustrated based on Experiment~$9$ summarized in Table \ref{table:RERS}.}
\label{Results_Trace}
\end{figure}

\subsubsection{Performance Evaluation}
Fig. \ref{Results_Trace} illustrates the trace of the location estimate $\boldsymbol{y}_{\textrm{RE}\&\textrm{RS}}$ of \eqref{LocEst_Filtering} drawn on the experimental site's floor plan. The solid black line represents the ground-truth user trajectory. 
Three benchmarks are considered. The first is LLS-RS without CDA \cite{Guvenc2008}. The second and third ones are the existing algorithms similar to CDA mentioned above, namely LMeS \cite{Qiao2014} and RWGH \cite{Chen1999}, which are representatives of the median- and mean-based estimators, respectively. 

Several key observations are made. First, it is verified that CDA effectively mitigates the NLoS effect by comparing LLS-RS and the remaining CDA-based techniques. Second, our tandem filtering method outperforms LMeS and RWGH at most MPs, confirming that the remaining PELs are more reliable than the filtered ones. Third, a considerable error is observed at a few MPs, e.g., MP $18$. Note that our filtering operation hinges on a relative comparison between given PELs. It always returns $12$ PELs according to the current configuration. In other words, a few unreliable PELs exposed to severe NLoS environments can remain, causing a significant positioning error. It is essential to incorporate the previous location estimate and IMU measurements, which are dealt with in the following subsection.

  \subsection{CDA-Based KF Integration with PDR}\label{subsection:Cov_Update} 

\subsubsection{Algorithm Description}
The goal of this subsection is to extend CDA's usage into PDR. Consider that the previous location estimate is initially given as $\boldsymbol{y}_0$, and the user moves to the targeted MP. His mobility pattern can be detectable by IMUs' measurements, expressed as a linear piecewise $\boldsymbol{v}=[v_x, v_y]^T$ where $v_x$ and $v_y$ are projected movements onto the directions of $x$ and $y$ axes, respectively \cite{Han2021}. Given $\boldsymbol{y}_0$, $\boldsymbol{v}$, and the current PELs $\mathcal{Z}$ specified in \eqref{PEL_set},  we aim at finding the current location estimate.

\begin{figure}
\centering 
{\includegraphics[width=8.8cm]{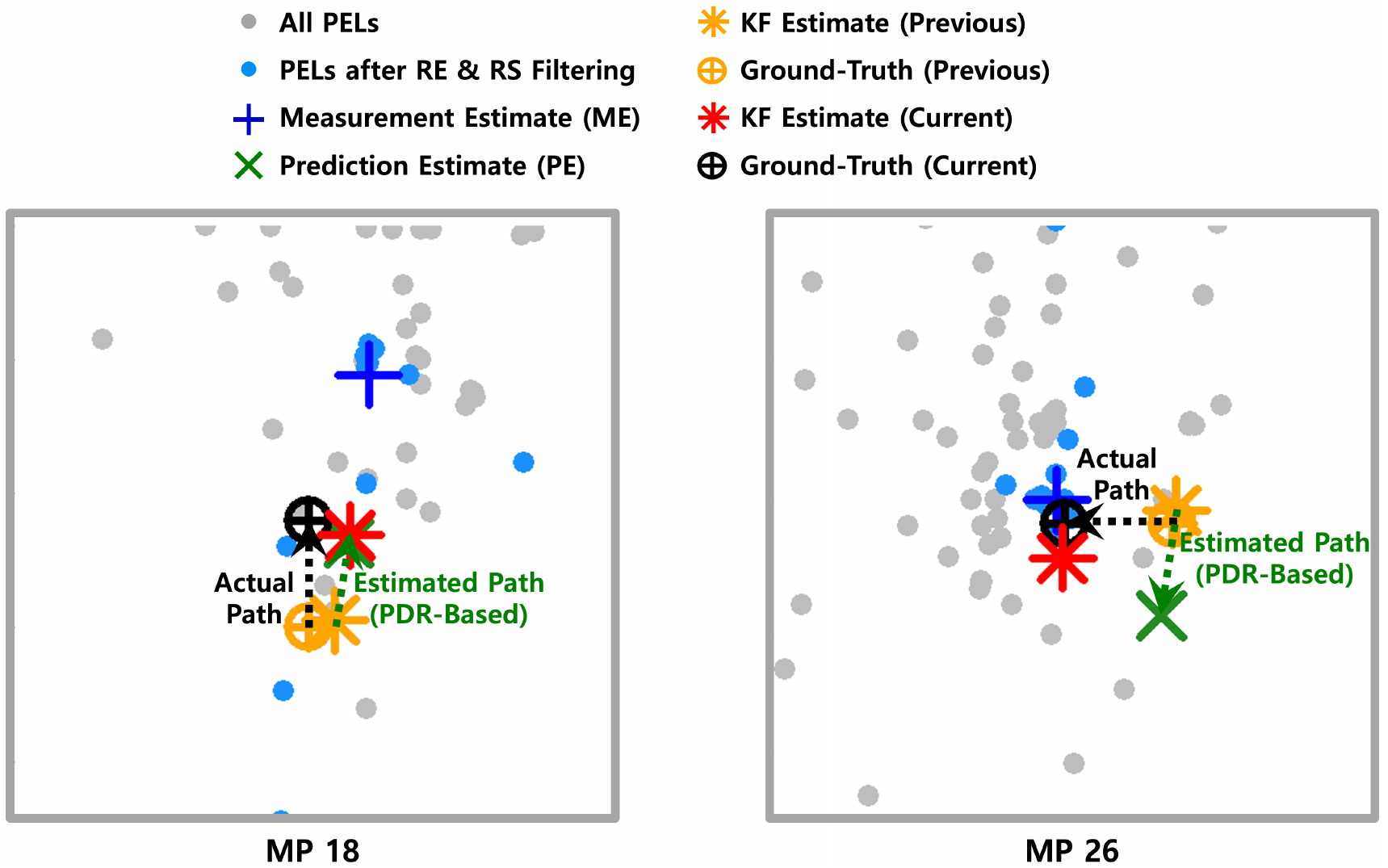}} 
\caption{Graphical illustration of KF using CDA-based covariance update specified in \eqref{R_Update} and \eqref{Q_Update}. The left (right) subfigure,  the result at MP $18$ ($26$) in Experiment $9$, shows the example when ME (PE) is far from the ground truth. It is observed that the proposed KF  can find the location estimate closer to the ground truth for both cases by properly weighing between  ME and PE.}
\label{KF_Example}
\end{figure}

Two types of estimates are derived to this end. First, recall $\boldsymbol{y}_{\textrm{RE}\&\textrm{RS}}$ specified in~\eqref{LocEst_Filtering}, which only uses the current measurement $\mathcal{Z}$. We call it a \emph{measurement estimate} (ME).  Second, the current location can be {predicted} by adding $\boldsymbol{v}$ to $\boldsymbol{y}_0$, say ${\boldsymbol{y}}_0+\boldsymbol{v}$. We call it a \emph{prediction estimate} (PE). One typical technique combining the two is a KF, which is optimal when the measurement and prediction errors are Gaussian processes, and the two covariance matrices are known. Specifically, a new position estimate, denoted by $\boldsymbol{y}_{\textrm{KF}}$, can be derived  by a weighted linear combination of PE and ME, given as 
\begin{align}\label{KF}
\boldsymbol{y}_{\textrm{KF}}= \mathbf{G}\underbrace{\boldsymbol{y}_{\textrm{RE}\&\textrm{RS}}}_{\textrm{ME}}+(\boldsymbol{I}_2-\mathbf{G})(\underbrace{{\boldsymbol{y}}_0+\boldsymbol{v}}_{\textrm{PE}}),
\end{align}
where $\boldsymbol{I}_2$ is $2$-by-$2$ identity matrix and $\boldsymbol{G}\in \mathbb{R}^{2 \times 2}$ represents the weight to ME, called a Kalman gain. The detailed derivation of $\boldsymbol{G}$ is summarized in Appendix C.

Note that we can compute $\boldsymbol{G}$ when ME and PE's error covariance matrices are given, denoted by $\boldsymbol{R}$ and $\boldsymbol{Q}$, respectively.
To the final estimate be accurate, it requires updating $\boldsymbol{R}$ and $\boldsymbol{Q}$ reflecting on current ME and PEs' noisy levels, yet challenging due to the lack of sufficient statistics.  One conventional way is to use their long-term historical beliefs without considering the current situation, limiting the KF performance~\cite{Zhang2020}.

Assuming that the projected errors onto $x$- and $y$-axes are independent for both ME and PE, we propose a novel CDA-based real-time covariance update as follows. 
\begin{itemize}
\item \emph{Measurement Covariance Matrix}: The spatial variance of the PELs in $\mathcal{Z}_{\textrm{RE}\&\textrm{RS}}$ of \eqref{LocEst_Filtering} is used as the diagonal terms of $\boldsymbol{R}$, namely,
\begin{align}\label{R_Update}
\boldsymbol{R}=\mathsf{diag}[\mathsf{var}(\boldsymbol{z}_{\ell}(1)), \mathsf{var}(\boldsymbol{z}_{\ell}(2))], \quad \boldsymbol{z}_{\ell}\in\mathcal{Z}_{\textrm{RE}\&\textrm{RS}}, 
\end{align}
where $\boldsymbol{z}_{\ell}(1)$ and $\boldsymbol{z}_{\ell}(2)$ are $x$ and $y$ coordinates of PEL~$\ell$. 
\item \emph{Prediction Covariance Matrix}: Given $\boldsymbol{y}_0$, the variance of PE is equivalent to the variance of the mobility pattern $\boldsymbol{v}$, expressed as 
\begin{align}\label{Q_Update}
\boldsymbol{Q}=\mathsf{diag}[\mathsf{var}(v_x), \mathsf{var}(v_y)], 
\end{align}
where each term can be computed from a sequence of IMU measurements when moving from previous and current MPs.\footnote{Specifically, the terms $v_x$ and $v_y$ are respectively expressed as $d \cos(\theta)$ and $d \sin(\theta)$, where $d$ is the moving distance, and $\theta$ is the heading direction. The smartphone's accelerometer and gyroscope enable us to keep measuring the former and latter, respectively. With their aids, we can compute the variances of $v_x$ and $v_y$ in real time. }
\end{itemize}

\begin{remark}[When KF Meets CDA]\emph{CDA makes KF more effective when deriving the measurement covariance matrix $\boldsymbol{R}$ to compute the Kalman gain $\boldsymbol{G}$ in \eqref{KF}.  Fig. \ref{KF_Example} illustrates two KF examples when either ME or RE has a significant positioning error. One shows that the final estimate $\boldsymbol{y}_{\textrm{KF}}$ is placed nearer to the ground truth than both ME and RE.  Primarily,  $\boldsymbol{R}$ in \eqref{R_Update} can capture the current RTTs' noisy level from PELs' spatial distribution. For example, as PELs are more dispersed (concentrated) with higher (smaller) spatial covariance, it can be interpreted that the current RTTs are more (less) corrupted. Besides, the tandem filter in Sec. \ref{subsection:PEL_Filter} excludes PELs severely biased. Thus, the remaining PELs follow a Gaussian distribution without bias to a particular direction. As a result, the derived $\boldsymbol{R}$ fulfills the prerequisites for optimal KF mentioned~above.}
\end{remark}

\begin{table*}
\centering
\caption{Summary of the experimental results in Sec. \ref{Sec3}. }\label{table:RERS}
\begin{tabular}{ll|cccccccccccc|c}
\toprule
\multicolumn{2}{c|}{{Experiment \#}}                                                                                                         & $1$             & $2$    & $3$    & $4$    & $5$    & $6$    & $7$    & $8$    & $9$    & $10$   & $11$   & $12$   & {Total} \\ \midrule
\multicolumn{1}{c|}{\multirow{2}{*}{{LLS-RS \cite{Guvenc2008}}}}                                                                                  & $\mathsf{avg}$ (m) & $6.24$ & $7.17$ & $6.80$ & $7.07$ & $6.08$ & $6.18$ & $6.29$ & $7.27$ & $6.58$ & $7.52$ & $6.54$ & $6.52$ & $6.69$           \\ 
\multicolumn{1}{c|}{}                                                                                                                  & $\mathsf{std}$ (m) & 3.08          & $4.31$ & $4.33$ & $4.62$ & $4.15$ & $3.01$ & $4.18$ & $5.21$ & $3.72$ & $5.67$ & $4.49$ & $3.37$ & $4.18$           \\ \midrule
\multicolumn{1}{c|}{\multirow{2}{*}{{CDA only}}}                                                                              & $\mathsf{avg}$ (m) & $1.80$          & $1.90$ & $2.14$ & $1.77$ & $2.33$ & $1.64$ & $1.76$ & $1.69$ & $1.53$ & $1.56$ & $1.44$ & $1.71$ & $1.77$           \\ 
\multicolumn{1}{c|}{}                                                                                                                  & $\mathsf{std}$ (m) & $1.35$          & $1.52$ & $1.83$ & $1.51$ & $2.07$ & $1.27$ & $1.22$ & $1.28$ & $1.04$ & $1.34$ & $0.96$ & $1.24$ & $1.39$           \\ \midrule
\multicolumn{1}{c|}{\multirow{2}{*}{{\begin{tabular}[c]{@{}c@{}}CDA \& PDR\\ (KF w/ deterministic $\boldsymbol{R}$ \& $\boldsymbol{Q}$)\end{tabular}}}} & $\mathsf{avg}$ (m) & $1.73$          & $1.91$ & $1.98$ & $1.80$ & $2.18$ & $1.54$ & $1.54$ & $1.75$ & $1.41$ & $1.41$ & $1.38$ & $1.68$ & $1.69$           \\ 
\multicolumn{1}{c|}{}                                                                                                                  & $\mathsf{std}$ (m) & $1.29$          & $1.30$ & $1.64$ & $1.31$ & $1.84$ & $1.11$ & $1.09$ & $1.13$ & $0.82$ & $1.03$ & $0.86$ & $1.06$ & $1.21$           \\ \midrule
\multicolumn{1}{c|}{\multirow{2}{*}{{\begin{tabular}[c]{@{}c@{}}CDA \& PDR\\ (KF w/ updating $\boldsymbol{R}$ \eqref{R_Update} \& $\boldsymbol{Q}$ \eqref{Q_Update})\end{tabular}}}}     & $\mathsf{avg}$ (m) & $1.59$          & $2.11$ & $2.06$ & $1.86$ & $2.26$ & $1.36$ & $1.38$ & $1.62$ & $1.25$ & $1.24$ & $1.45$ & $1.63$ & $1.65$           \\ 
\multicolumn{1}{c|}{}                                                                                                                  & $\mathsf{std}$ (m) & $1.04$          & $1.27$ & $1.40$ & $1.19$ & $1.44$ & $0.98$ & $0.88$ & $0.93$ & $0.65$ & $0.79$ & $0.69$ & $0.89$ & $1.01$           \\ \bottomrule
\end{tabular}\end{table*}

\begin{figure}
\centering 
{\includegraphics[width=9.0cm]{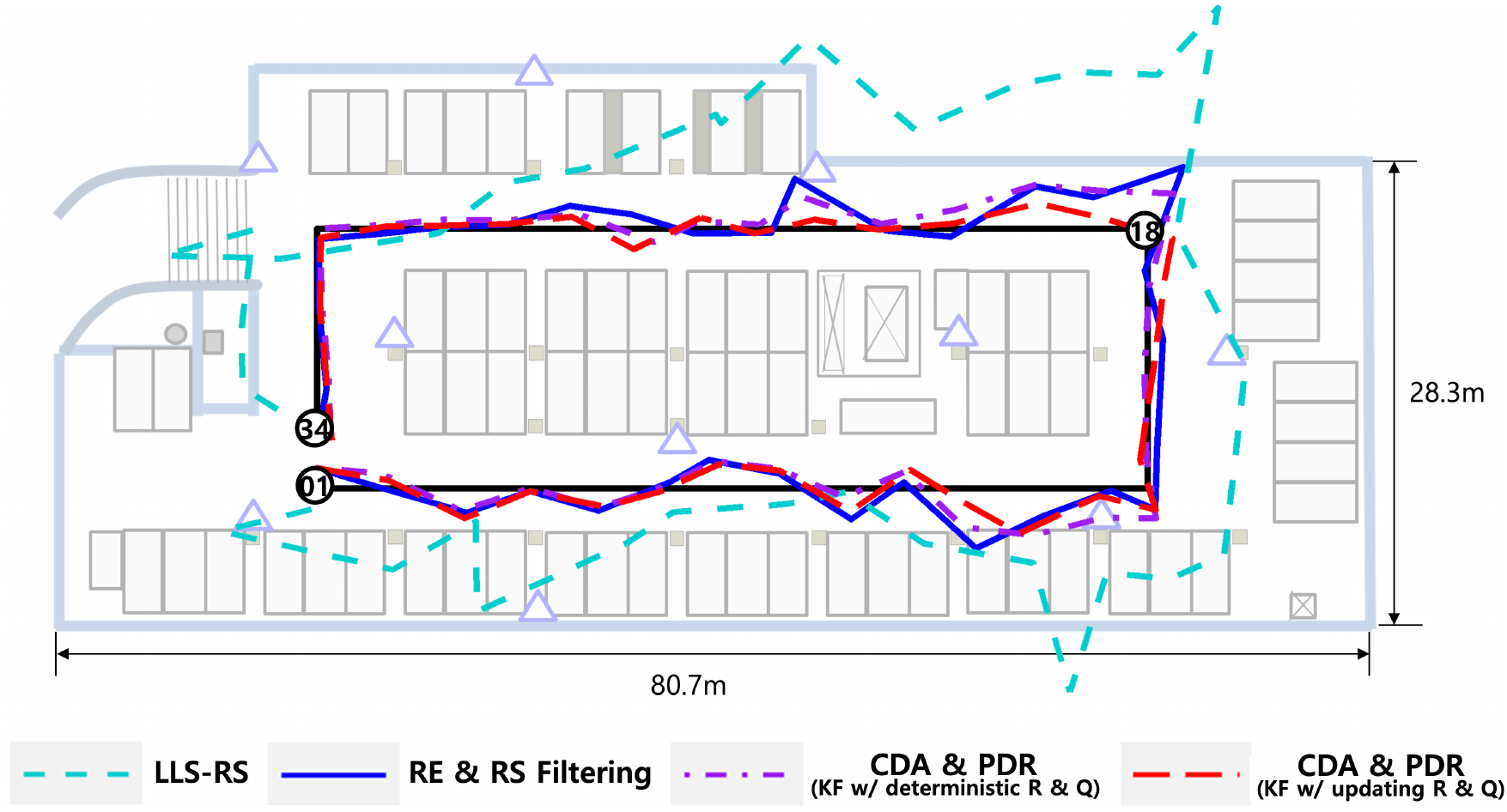}} 
\caption{Graphical representation of location traces when CDA-based PDR are used with KF, drawn on the floor plan of the experimental site. All results are illustrated based on Experiment~$9$ summarized in Table~\ref{table:RERS}.}
\label{Results_Trace_w_PDR}
\end{figure}

\subsubsection{Performance Evaluation}
Fig. \ref{Results_Trace_w_PDR} illustrates the trace of positioning results when PDR is incorporated with CDA with the proposed real-time covariance updates of \eqref{R_Update} and \eqref{Q_Update}. As benchmarks, we consider RE \& RS filtering explained in Sec. \ref{subsection:PEL_Filter} and PDR with CDA using 
identity matrices as the covariance matrices. It is shown that the proposed approach's positioning result is relatively accurate and stable for most MPs. Recall MP $18$ in which the positioning errors of ME $\boldsymbol{y}_{\textrm{RE}\&\textrm{RS}}$ remain significant. KF with CDA attempts to find the user's location by properly adjusting ME and PE's weights, estimating the user's location closer to the ground truth.

Table~\ref{table:RERS} summarizes absolute position errors' \emph{average} ($\mathsf{avg}$) and \emph{standard deviation} ($\mathsf{std}$) at each MP of all $12$ experiments. The examples in Figs. \ref{Results_Trace}, \ref{KF_Example}, and \ref{Results_Trace_w_PDR} are based on Experiment $9$'s results. Several interesting observations are made. First, all CDA-driven approaches always outperform the conventional GP without CDA, i.e., LLR-RS \cite{Guvenc2008}. Second, the $\mathsf{avg}$ of the CDA \& PDR approach is less than that of RE \& RS filtering (i.e., standalone CDA without PDR) in most experiments. In contrast, the opposite is also observed in a few ones (e.g., Experiment $4$) when IMU measurements are severely noisy. Third, contrary to $\mathsf{avg}$, the CDA \& PDR's $\mathsf{std}$ is always smaller than the standalone CDA approach. Last, the significant gain on reducing std is observed when the proposed updates of covariance matrices $\boldsymbol{R}$ and $\boldsymbol{Q}$, verifying CDA to provide not only precise but also stable positioning results.  

We conducted another field experiment at a different site (Geumnamno4-ga subway station B3F, Gwangju, South Korea), observing similar results to the above experiment. For brevity, we summarize the results in Appendix D.

\section{Combinatorial Data Augmentation Helps Fingerprint-Based Positioning}\label{sec4}

This section revisits Problem \ref{fingerprint_Def}, exploiting the position estimates in Sec. \ref{Sec3} to tackle the issue of FBP without the ground-truth label.  The detailed algorithms are first elaborated, and their positioning results are compared.

\subsection{Algorithm Description}\label{subsection:FBPalgorithm}

This subsection explains our algorithm design, including input features \& data labeling, ML model, and training \& testing. 

\subsubsection{Input Features \& Data Labeling} Recall that we have the results of $12$ experiments, each of which comprises $34$ MPs, i.e., $|\mathcal{K}|=K=34$. In other words, we have $12 \times 34=408$ data samples. We denote the data sample's index $e\in \mathcal{S}$, where $\mathcal{S}$ is the set of data samples. 
We exclusively partition them into two parts; $70\%$ and $30\%$ samples are randomly selected for training and test, denoted by $\mathcal{S}_{\textrm{train}}$ and $\mathcal{S}_{\textrm{test}}$, respectively.

We consider three types of input features as follows. 
\begin{itemize} 
\item First, given the number of APs $N=10$, we use raw RTTs, denoted by $\boldsymbol{s}^{(e)}_1\in \mathbb{R}^{{10}\times 1}$, namely, 
\begin{align}\label{InputFeature1}
\boldsymbol{s}^{(e)}_1=\boldsymbol{\tau}^{(e)}, \quad   e\in \mathcal{S}.
\end{align} 
\item Second, given $120$ PELs defined on 2D coordinates, we concatenate all elements as input features, denoted by $\boldsymbol{s}^{(e)}_2\in \mathbb{R}^{240\times 1}$, namely,
\begin{align}\label{InputFeature2}
\boldsymbol{s}^{(e)}_2=\mathsf{cat}(\mathcal{Z}^{(e)}), \quad  e\in \mathcal{S},
\end{align}
where the PEL set $\mathcal{Z}$ is specified in \eqref{PEL_set} and $\mathsf{cat}(\cdot)$ means the operation of concatenate arrays. 
\item Third, given $12$ remaining PELs after RE \& RS filtering, we concatenate them as input features, denoted by $\boldsymbol{s}^{(e)}_3\in \mathbb{R}^{24\times 1}$, namely,
\begin{align}\label{InputFeature3}
\boldsymbol{s}^{(e)}_3=\mathsf{cat}(\mathcal{Z}_{\textrm{RE}\&\textrm{RS}}^{(e)}), \quad  e\in \mathcal{S},
\end{align}
where the remaining PEL set $\mathcal{Z}_{\textrm{RE}\&\textrm{RS}}$ is specified in \eqref{LocEst_Filtering}.
\end{itemize}

Next, we consider that the location estimates $\boldsymbol{y}_{\textrm{RE\&RS}}^{(e)}$ and $\boldsymbol{y}_{\textrm{KF}}^{(e)}$ specified in \eqref{LocEst_Filtering} and \eqref{KF} are used as data labels for the corresponding input features.  
For notational brevity, we simply express them as $\boldsymbol{y}_{j}^{(e)}\in \mathbb{R}^{2\times 1}$, where $j=\{1,2\}$ represent the former and latter, respectively. Last, we additionally consider a ground-truth location $\boldsymbol{x}$  as an ideal benchmark giving the achievable bound, denoted by $\boldsymbol{y}_0^{(e)}$.

\subsubsection{ML Models}\label{subsubsection:ML}

For FBP to be efficient, it is vital to choose an apt ML model that is operable under the two practical conditions stated below. First, we need to train an ML model fast before it is out-of-date due to slight environmental changes.
Second, we rely on users' data collection. The training data samples are thus limited, resulting in the resultant ML model being overfitted.  
In those senses, we consider the following three ML models. 
\begin{itemize}
\item SVR is a regression version of the support vector machine less affected by a few outliers. Let $i=\{1, 2, 3\}$ indicate the type of input features explained above.
We define $\boldsymbol{w}_i^T\boldsymbol{s}_i^{(e)}+\boldsymbol{b}_i$ as the prediction of the input $\boldsymbol{s}_i^{(e)}$, where $\boldsymbol{w}_i\in \mathbb{R}^{2\times\mathsf{dim}(\boldsymbol{s}_i^{(e)})}$ and $\boldsymbol{b}\in \mathbb{R}^{2\times 1}$ are weight and bias, respectively. Given the data label $\boldsymbol{y}_j^{(e)}$, the problem is formulated as 
\begin{align}
&\min_{\boldsymbol{w}_i, \boldsymbol{b}_i, \{\xi^{(e)}\}} \frac{1}{2}\|\boldsymbol{w}_i\|_2^2+c \sum_{e\in\mathcal{S}} |\xi^{(e)}|\nonumber\\
 \textrm{s.t.} &\quad  \|\boldsymbol{y}_j^{(e)}-\boldsymbol{w}_i^T\boldsymbol{s}_i^{(e)}-\boldsymbol{b}_i\|_2\leq \epsilon+ |\xi^{(e)}|,\quad e\in \mathcal{S}_{\textrm{train}} \nonumber. 
\end{align}
The first and second terms in the objective function represent the regularization penalty and the prediction penalty with an error more significant than the threshold $\epsilon$, respectively. We set $\epsilon=0.1$ as a default value. The hyper-parameter $c$ is a weight to prediction error. Besides, we use a \emph{radial basis function} (RBF) with parameter $\gamma$ for a kernel trick addressing a nonlinear prediction. The parameters $c$ and $\gamma$ are exhaustedly optimized by choosing the best pair among various parameter settings.  
\item \emph{Random Forest} (RF) is an ensemble ML method efficient to avoid overfitting due to the limited number of training data. Specifically, we randomly make $500$ independent feature subsets. Each subset includes one-third of the full features.  Then, each tree is optimized by selecting a few features among the corresponding subset to minimize the uncertainty of its regression outcome, quantified by an entropy. The final estimate is computed by averaging all decision trees' results. 
\item \emph{Deep Neural Network} (DNN) is the state-of-the-art ML technique widely used in many applications.  For that reason, we use DNN as a benchmark though it does not meet the two conditions mentioned above. We consider a fully-connected network with one hidden layer including $\frac{\mathsf{dim}(\boldsymbol{s}_i^{(e)})+2}{2}$ nodes, where the number $2$ in the numerator represents the output's dimension, equivalent to the dimension of the location estimate. We use Adam optimizer with a single batch. The early stopping rule is adopted with the maximum number of epochs being $1000$.
\end{itemize}
\subsubsection{Training \& Test} Given $\mathcal{S}_{\textrm{train}}$ and $\mathcal{S}_{\textrm{test}}$, we train and test each ML model using different pairs of input features and data labels. We repeat the process five times by selecting different $\mathcal{S}_{\textrm{train}}$ and $\mathcal{S}_{\textrm{test}}$.  

\begin{figure*}[t]
\centering
\includegraphics[width=18cm]{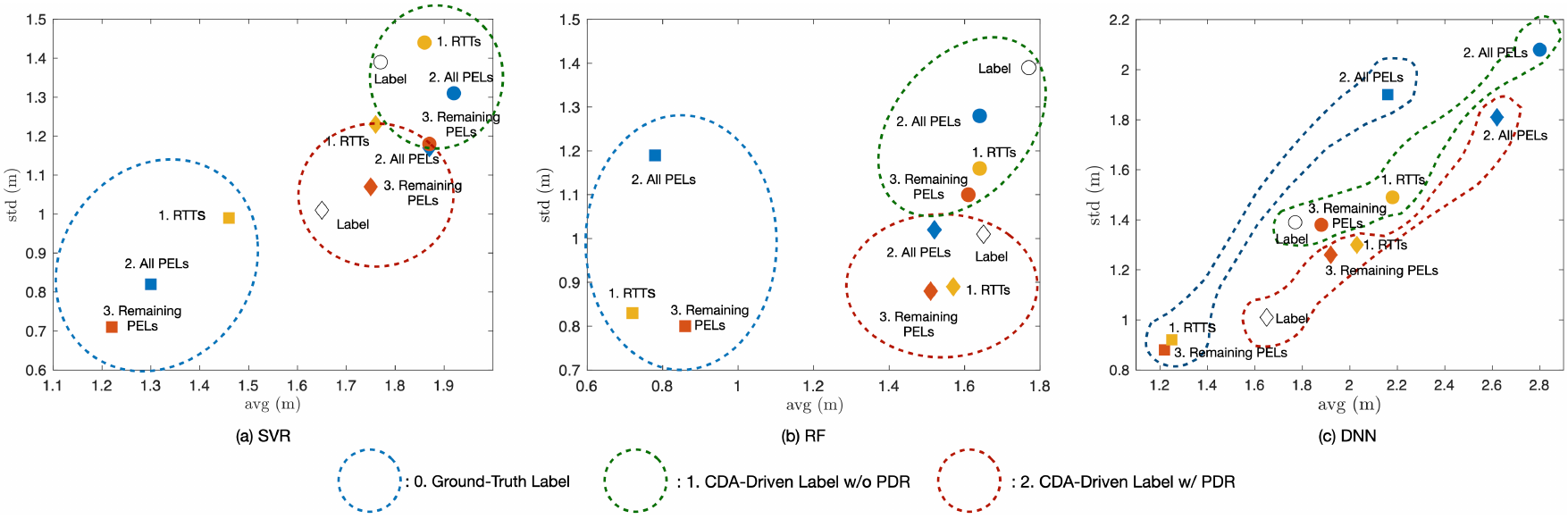}
\caption{FBP algorithms' $\mathsf{avg}$ \& $\mathsf{std}$ with various combinations of features \& labels when different ML models are used. } \label{FBP_Graph}
\end{figure*}

\begin{table*}
\centering
\caption{Summary of the experimental results in Sec. \ref{sec4}. }\label{Table:FBP}
\resizebox{\textwidth}{!}{%
\begin{tabular}{ccc|cccccc|cccccc|cccccc}
\toprule
\multicolumn{1}{c|}{\multirow{2}{*}{ML Models}} & \multicolumn{1}{c|}{\multirow{2}{*}{Input Features}}                                                & \multirow{2}{*}{Metric} & \multicolumn{6}{c|}{\multirow{2}{*}{0. Ground-truth Label}}                                                                                       & \multicolumn{6}{c|}{\multirow{2}{*}{1. CDA-based Label (w/o PDR)}}                                                                                & \multicolumn{6}{c}{\multirow{2}{*}{2. CDA-based label (w/ PDR)}}                                                                                 \\
\multicolumn{1}{c|}{}                           & \multicolumn{1}{c|}{}                                                                               &                         & \multicolumn{6}{c|}{}                                                                                                                             & \multicolumn{6}{c|}{}                                                                                                                             & \multicolumn{6}{c}{}                                                                                                                             \\ \midrule
\multicolumn{3}{c|}{Experiment \#}                                                                                                                                              & \multicolumn{1}{c|}{1}    & \multicolumn{1}{c|}{2}    & \multicolumn{1}{c|}{3}    & \multicolumn{1}{c|}{4}    & \multicolumn{1}{c|}{5}    & Total & \multicolumn{1}{c|}{1}    & \multicolumn{1}{c|}{2}    & \multicolumn{1}{c|}{3}    & \multicolumn{1}{c|}{4}    & \multicolumn{1}{c|}{5}    & Total & \multicolumn{1}{c|}{1}    & \multicolumn{1}{c|}{2}    & \multicolumn{1}{c|}{3}    & \multicolumn{1}{c|}{4}    & \multicolumn{1}{c|}{5}    & Total \\ \midrule
\multicolumn{1}{c|}{\multirow{6}{*}{SVR}}       & \multicolumn{1}{c|}{\multirow{2}{*}{1. Raw RTTs}}                                                        & $\mathsf{avg}$ (m)                 & \multicolumn{1}{c|}{$1.44$} & \multicolumn{1}{c|}{$1.42$} & \multicolumn{1}{c|}{$1.49$} & \multicolumn{1}{c|}{$1.5$}  & \multicolumn{1}{c|}{$1.47$} & $1.46$  & \multicolumn{1}{c|}{$1.81$} & \multicolumn{1}{c|}{$1.68$} & \multicolumn{1}{c|}{$2.03$} & \multicolumn{1}{c|}{$1.85$} & \multicolumn{1}{c|}{$1.92$} & $1.86$  & \multicolumn{1}{c|}{$1.71$} & \multicolumn{1}{c|}{$1.54$} & \multicolumn{1}{c|}{$1.9$}  & \multicolumn{1}{c|}{$1.82$} & \multicolumn{1}{c|}{$1.84$} & $1.76$  \\ 
\multicolumn{1}{c|}{}                           & \multicolumn{1}{c|}{}                                                                               & $\mathsf{std}$ (m)                 & \multicolumn{1}{c|}{$0.98$} & \multicolumn{1}{c|}{$0.83$} & \multicolumn{1}{c|}{$1.07$} & \multicolumn{1}{c|}{$0.95$} & \multicolumn{1}{c|}{$1.12$} & $0.99$  & \multicolumn{1}{c|}{$1.37$} & \multicolumn{1}{c|}{$1.14$} & \multicolumn{1}{c|}{$1.76$} & \multicolumn{1}{c|}{$1.32$} & \multicolumn{1}{c|}{$1.6$}  & $1.44$  & \multicolumn{1}{c|}{$1.23$} & \multicolumn{1}{c|}{$0.87$} & \multicolumn{1}{c|}{$1.36$} & \multicolumn{1}{c|}{$1.16$} & \multicolumn{1}{c|}{$1.53$} & $1.23$  \\ \cline{2-21} 
\multicolumn{1}{c|}{}                           & \multicolumn{1}{c|}{\multirow{2}{*}{2. All PELs}}                                                   & $\mathsf{avg}$ (m)                 & \multicolumn{1}{c|}{$1.36$} & \multicolumn{1}{c|}{$1.21$} & \multicolumn{1}{c|}{$1.26$} & \multicolumn{1}{c|}{$1.31$} & \multicolumn{1}{c|}{$1.35$} & $1.3$   & \multicolumn{1}{c|}{$1.97$} & \multicolumn{1}{c|}{$1.96$} & \multicolumn{1}{c|}{$1.94$} & \multicolumn{1}{c|}{$1.97$} & \multicolumn{1}{c|}{$1.75$} & $1.92$  & \multicolumn{1}{c|}{$1.85$} & \multicolumn{1}{c|}{$1.73$} & \multicolumn{1}{c|}{$1.94$} & \multicolumn{1}{c|}{$1.93$} & \multicolumn{1}{c|}{$1.89$} & $1.87$  \\ 
\multicolumn{1}{c|}{}                           & \multicolumn{1}{c|}{}                                                                               & $\mathsf{std}$ (m)                 & \multicolumn{1}{c|}{$0.76$} & \multicolumn{1}{c|}{$0.66$} & \multicolumn{1}{c|}{$0.77$} & \multicolumn{1}{c|}{$0.8$}  & \multicolumn{1}{c|}{$1.09$} & $0.82$  & \multicolumn{1}{c|}{$1.3$}  & \multicolumn{1}{c|}{$1.4$}  & \multicolumn{1}{c|}{$1.39$} & \multicolumn{1}{c|}{$1.19$} & \multicolumn{1}{c|}{$1.28$} & $1.31$  & \multicolumn{1}{c|}{$1.26$} & \multicolumn{1}{c|}{$1.07$} & \multicolumn{1}{c|}{$1.21$} & \multicolumn{1}{c|}{$1.13$} & \multicolumn{1}{c|}{$1.2$}  & $1.17$  \\ \cline{2-21} 
\multicolumn{1}{c|}{}                           & \multicolumn{1}{c|}{\multirow{2}{*}{\begin{tabular}[c]{@{}c@{}}3. Remaining  \\ PELs\end{tabular}}} & $\mathsf{avg}$ (m)                 & \multicolumn{1}{c|}{$1.28$} & \multicolumn{1}{c|}{$1.22$} & \multicolumn{1}{c|}{$1.23$} & \multicolumn{1}{c|}{$1.2$}  & \multicolumn{1}{c|}{$1.15$} & $1.22$  & \multicolumn{1}{c|}{$1.8$}  & \multicolumn{1}{c|}{$2.02$} & \multicolumn{1}{c|}{$1.81$} & \multicolumn{1}{c|}{$1.93$} & \multicolumn{1}{c|}{$1.78$} & $1.87$  & \multicolumn{1}{c|}{$1.67$} & \multicolumn{1}{c|}{$1.83$} & \multicolumn{1}{c|}{$1.77$} & \multicolumn{1}{c|}{$1.77$} & \multicolumn{1}{c|}{$1.7$}  & $1.75$  \\ 
\multicolumn{1}{c|}{}                           & \multicolumn{1}{c|}{}                                                                               & $\mathsf{std}$ (m)                 & \multicolumn{1}{c|}{$0.78$} & \multicolumn{1}{c|}{$0.72$} & \multicolumn{1}{c|}{$0.71$} & \multicolumn{1}{c|}{$0.73$} & \multicolumn{1}{c|}{$0.6$}  & $0.71$  & \multicolumn{1}{c|}{$1.2$}  & \multicolumn{1}{c|}{$1.27$} & \multicolumn{1}{c|}{$1.12$} & \multicolumn{1}{c|}{$1.18$} & \multicolumn{1}{c|}{$1.14$} & $1.18$  & \multicolumn{1}{c|}{$1.17$} & \multicolumn{1}{c|}{$1.05$} & \multicolumn{1}{c|}{$1.1$}  & \multicolumn{1}{c|}{$1.07$} & \multicolumn{1}{c|}{$0.98$} & $1.07$  \\ \midrule
\multicolumn{1}{c|}{\multirow{6}{*}{RF}}        & \multicolumn{1}{c|}{\multirow{2}{*}{1. Raw RTTs}}                                                        & $\mathsf{avg}$ (m)                 & \multicolumn{1}{c|}{$0.68$} & \multicolumn{1}{c|}{$0.68$} & \multicolumn{1}{c|}{$0.71$} & \multicolumn{1}{c|}{$0.8$}  & \multicolumn{1}{c|}{$0.73$} & $0.72$  & \multicolumn{1}{c|}{$1.73$} & \multicolumn{1}{c|}{$1.63$} & \multicolumn{1}{c|}{$1.63$} & \multicolumn{1}{c|}{$1.65$} & \multicolumn{1}{c|}{$1.57$} & $1.64$  & \multicolumn{1}{c|}{$1.61$} & \multicolumn{1}{c|}{$1.49$} & \multicolumn{1}{c|}{$1.51$} & \multicolumn{1}{c|}{$1.65$} & \multicolumn{1}{c|}{$1.58$} & $1.57$  \\ 
\multicolumn{1}{c|}{}                           & \multicolumn{1}{c|}{}                                                                               & $\mathsf{std}$ (m)                 & \multicolumn{1}{c|}{$0.84$} & \multicolumn{1}{c|}{$0.74$} & \multicolumn{1}{c|}{$0.76$} & \multicolumn{1}{c|}{$0.96$} & \multicolumn{1}{c|}{$0.84$} & $0.83$  & \multicolumn{1}{c|}{$1.24$} & \multicolumn{1}{c|}{$1.14$} & \multicolumn{1}{c|}{$1.18$} & \multicolumn{1}{c|}{$1.05$} & \multicolumn{1}{c|}{$1.2$}  & $1.16$  & \multicolumn{1}{c|}{$0.96$} & \multicolumn{1}{c|}{$0.82$} & \multicolumn{1}{c|}{$0.91$} & \multicolumn{1}{c|}{$0.86$} & \multicolumn{1}{c|}{$0.92$} & $0.89$  \\ \cline{2-21} 
\multicolumn{1}{c|}{}                           & \multicolumn{1}{c|}{\multirow{2}{*}{2. All PELs}}                                                   & $\mathsf{avg}$ (m)                 & \multicolumn{1}{c|}{$0.64$} & \multicolumn{1}{c|}{$0.7$}  & \multicolumn{1}{c|}{$0.66$} & \multicolumn{1}{c|}{$1.01$} & \multicolumn{1}{c|}{$0.9$}  & $0.78$  & \multicolumn{1}{c|}{$1.67$} & \multicolumn{1}{c|}{$1.55$} & \multicolumn{1}{c|}{$1.7$}  & \multicolumn{1}{c|}{$1.61$} & \multicolumn{1}{c|}{$1.68$} & $1.64$  & \multicolumn{1}{c|}{$1.45$} & \multicolumn{1}{c|}{$1.42$} & \multicolumn{1}{c|}{$1.53$} & \multicolumn{1}{c|}{$1.61$} & \multicolumn{1}{c|}{$1.59$} & $1.52$  \\ 
\multicolumn{1}{c|}{}                           & \multicolumn{1}{c|}{}                                                                               & $\mathsf{std}$ (m)                 & \multicolumn{1}{c|}{$0.81$} & \multicolumn{1}{c|}{$1.14$} & \multicolumn{1}{c|}{$1.03$} & \multicolumn{1}{c|}{$1.41$} & \multicolumn{1}{c|}{$1.58$} & $1.19$  & \multicolumn{1}{c|}{$1.27$} & \multicolumn{1}{c|}{$1.08$} & \multicolumn{1}{c|}{$1.59$} & \multicolumn{1}{c|}{$1.21$} & \multicolumn{1}{c|}{$1.27$} & $1.28$  & \multicolumn{1}{c|}{$0.81$} & \multicolumn{1}{c|}{$0.85$} & \multicolumn{1}{c|}{$1.17$} & \multicolumn{1}{c|}{$1.09$} & \multicolumn{1}{c|}{$1.17$} & $1.02$  \\ \cline{2-21} 
\multicolumn{1}{c|}{}                           & \multicolumn{1}{c|}{\multirow{2}{*}{\begin{tabular}[c]{@{}c@{}}3. Remaining  \\ PELs\end{tabular}}} & $\mathsf{avg}$ (m)                 & \multicolumn{1}{c|}{$0.88$} & \multicolumn{1}{c|}{$0.77$} & \multicolumn{1}{c|}{$0.9$}  & \multicolumn{1}{c|}{$0.89$} & \multicolumn{1}{c|}{$0.84$} & $0.86$  & \multicolumn{1}{c|}{$1.57$} & \multicolumn{1}{c|}{$1.72$} & \multicolumn{1}{c|}{$1.54$} & \multicolumn{1}{c|}{$1.67$} & \multicolumn{1}{c|}{$1.56$} & $1.61$  & \multicolumn{1}{c|}{$1.51$} & \multicolumn{1}{c|}{$1.55$} & \multicolumn{1}{c|}{$1.42$} & \multicolumn{1}{c|}{$1.51$} & \multicolumn{1}{c|}{$1.54$} & $1.51$  \\ 
\multicolumn{1}{c|}{}                           & \multicolumn{1}{c|}{}                                                                               & $\mathsf{std}$ (m)                 & \multicolumn{1}{c|}{$0.75$} & \multicolumn{1}{c|}{$0.79$} & \multicolumn{1}{c|}{$0.89$} & \multicolumn{1}{c|}{$0.8$}  & \multicolumn{1}{c|}{$0.75$} & $0.8$   & \multicolumn{1}{c|}{$1.06$} & \multicolumn{1}{c|}{$1.17$} & \multicolumn{1}{c|}{$1.06$} & \multicolumn{1}{c|}{$1.06$} & \multicolumn{1}{c|}{$1.14$} & $1.1$   & \multicolumn{1}{c|}{$0.87$} & \multicolumn{1}{c|}{$0.88$} & \multicolumn{1}{c|}{$0.86$} & \multicolumn{1}{c|}{$0.88$} & \multicolumn{1}{c|}{$0.92$} & $0.88$  \\ \midrule
\multicolumn{1}{c|}{\multirow{6}{*}{DNN}}       & \multicolumn{1}{c|}{\multirow{2}{*}{1. Raw RTTs}}                                                        & $\mathsf{avg}$ (m)                 & \multicolumn{1}{c|}{$1.18$} & \multicolumn{1}{c|}{$1.19$} & \multicolumn{1}{c|}{$1.24$} & \multicolumn{1}{c|}{$1.23$} & \multicolumn{1}{c|}{$1.42$} & $1.25$  & \multicolumn{1}{c|}{$2.06$} & \multicolumn{1}{c|}{$2.18$} & \multicolumn{1}{c|}{$2.28$} & \multicolumn{1}{c|}{$2.1$}  & \multicolumn{1}{c|}{$2.3$}  & $2.18$  & \multicolumn{1}{c|}{$1.87$} & \multicolumn{1}{c|}{$1.73$} & \multicolumn{1}{c|}{$2.4$}  & \multicolumn{1}{c|}{$1.97$} & \multicolumn{1}{c|}{$2.19$} & $2.03$  \\ 
\multicolumn{1}{c|}{}                           & \multicolumn{1}{c|}{}                                                                               & $\mathsf{std}$ (m)                 & \multicolumn{1}{c|}{$0.96$} & \multicolumn{1}{c|}{$0.77$} & \multicolumn{1}{c|}{$0.86$} & \multicolumn{1}{c|}{$0.81$} & \multicolumn{1}{c|}{$1.2$}  & $0.92$  & \multicolumn{1}{c|}{$1.49$} & \multicolumn{1}{c|}{$1.31$} & \multicolumn{1}{c|}{$1.64$} & \multicolumn{1}{c|}{$1.27$} & \multicolumn{1}{c|}{$1.72$} & $1.49$  & \multicolumn{1}{c|}{$1.23$} & \multicolumn{1}{c|}{$0.89$} & \multicolumn{1}{c|}{$1.6$}  & \multicolumn{1}{c|}{$1.26$} & \multicolumn{1}{c|}{$1.5$}  & $1.3$   \\ \cline{2-21} 
\multicolumn{1}{c|}{}                           & \multicolumn{1}{c|}{\multirow{2}{*}{2. All PELs}}                                                   & $\mathsf{avg}$ (m)                 & \multicolumn{1}{c|}{$2.49$} & \multicolumn{1}{c|}{$1.5$}  & \multicolumn{1}{c|}{$2.01$} & \multicolumn{1}{c|}{$2.83$} & \multicolumn{1}{c|}{$1.95$} & $2.16$  & \multicolumn{1}{c|}{$2.72$} & \multicolumn{1}{c|}{$2.57$} & \multicolumn{1}{c|}{$2.98$} & \multicolumn{1}{c|}{$3$}    & \multicolumn{1}{c|}{$2.73$} & $2.8$   & \multicolumn{1}{c|}{$2.32$} & \multicolumn{1}{c|}{$2.71$} & \multicolumn{1}{c|}{$2.6$}  & \multicolumn{1}{c|}{$2.74$} & \multicolumn{1}{c|}{$2.74$} & $2.62$  \\ 
\multicolumn{1}{c|}{}                           & \multicolumn{1}{c|}{}                                                                               & $\mathsf{std}$ (m)                 & \multicolumn{1}{c|}{$2.18$} & \multicolumn{1}{c|}{$1.17$} & \multicolumn{1}{c|}{$2.06$} & \multicolumn{1}{c|}{$2.32$} & \multicolumn{1}{c|}{$1.77$} & $1.9$   & \multicolumn{1}{c|}{$2.08$} & \multicolumn{1}{c|}{$1.75$} & \multicolumn{1}{c|}{$2.42$} & \multicolumn{1}{c|}{$2.22$} & \multicolumn{1}{c|}{$1.95$} & $2.08$  & \multicolumn{1}{c|}{$1.46$} & \multicolumn{1}{c|}{$1.76$} & \multicolumn{1}{c|}{$1.66$} & \multicolumn{1}{c|}{$1.99$} & \multicolumn{1}{c|}{$2.19$} & $1.81$  \\ \cline{2-21} 
\multicolumn{1}{c|}{}                           & \multicolumn{1}{c|}{\multirow{2}{*}{\begin{tabular}[c]{@{}c@{}}3. Remaining  \\ PELs\end{tabular}}} & $\mathsf{avg}$ (m)                 & \multicolumn{1}{c|}{$1.31$} & \multicolumn{1}{c|}{$1.33$} & \multicolumn{1}{c|}{$1.09$} & \multicolumn{1}{c|}{$1.32$} & \multicolumn{1}{c|}{$1.06$} & $1.22$  & \multicolumn{1}{c|}{$1.92$} & \multicolumn{1}{c|}{$1.95$} & \multicolumn{1}{c|}{$1.72$} & \multicolumn{1}{c|}{$1.94$} & \multicolumn{1}{c|}{$1.85$} & $1.88$  & \multicolumn{1}{c|}{$1.99$} & \multicolumn{1}{c|}{2}    & \multicolumn{1}{c|}{$1.71$} & \multicolumn{1}{c|}{$2.09$} & \multicolumn{1}{c|}{$1.81$} & $1.92$  \\ 
\multicolumn{1}{c|}{}                           & \multicolumn{1}{c|}{}                                                                               & $\mathsf{std}$ (m)                 & \multicolumn{1}{c|}{$0.95$} & \multicolumn{1}{c|}{$0.91$} & \multicolumn{1}{c|}{$0.81$} & \multicolumn{1}{c|}{$1.09$} & \multicolumn{1}{c|}{$0.65$} & $0.88$  & \multicolumn{1}{c|}{$1.45$} & \multicolumn{1}{c|}{$1.47$} & \multicolumn{1}{c|}{$1.35$} & \multicolumn{1}{c|}{$1.31$} & \multicolumn{1}{c|}{$1.3$}  & $1.38$  & \multicolumn{1}{c|}{$1.34$} & \multicolumn{1}{c|}{$1.24$} & \multicolumn{1}{c|}{$1.32$} & \multicolumn{1}{c|}{$1.34$} & \multicolumn{1}{c|}{$1.07$} & $1.26$  \\ \bottomrule
\end{tabular}%
}
\end{table*}

\subsection{Performance Evaluation}

This subsection evaluates FBP when CDA-based input features and data labels are used. Fig. \ref{FBP_Graph} represents $\mathsf{avg}$ and $\mathsf{std}$ of FBP when various combinations of feature \& labels and different ML models are used. All numbers are derived by averaging $5$ experiments' results, each of which is summarized in Table \ref{Table:FBP}.  

Several interesting observations are made. First, it is shown that RF always provides smaller $\mathsf{avg}$ \& $\mathsf{std}$ of positioning error than the others for all feature-label pairs. On the other hand, DNN is the worst due to the limited number of training data samples, as expected in Sec. \ref{subsubsection:ML}. Second, when RF is used, both $\mathsf{avg}$ and $\mathsf{std}$ are less than those of the label (empty marks in each figure) for all feature-label pairs. 
The reason is that RF efficiently extracts essential yet unobservable factors from a few data samples, helping estimate the user's position in a harsh NLoS environment. 
Third, using the remaining PELs as input features [i.e., $\boldsymbol{s}^{(e)}_3$ of \eqref{InputFeature3}]  is shown to give the best performance in terms of $\mathsf{avg}$ \& $\mathsf{std}$, namely $1.61$ (m) and $1.01$ (m) 
with data label $\boldsymbol{y}_{1}^{(e)}$, and  $1.51$ (m) and $0.88$ (m) with data label $\boldsymbol{y}_{2}^{(e)}$. Especially, the input feature choice significantly reduces $\mathsf{std}$, which is near the FBP with the ground truth, say $0.80$ (m). In summary, the above observations confirm the CDA's effectiveness for FBP when the number of labeled data samples is limited.

\section{Concluding Remarks}\label{Sec6}

This paper has introduced CDA, a key enabler in integrating GP and DP for achieving real-time accurate and reliable positioning results.  CDA uses a low-complexity GP algorithm to generate many new data samples, transforming the positioning problems into data-domain issues, such as data filtering and fusion. Besides, CDA offers new input features and data labels for FBP without the ground-truth label, making DP practical. We have verified the effectiveness of CDA through field experiments with WiFi and IMUs. 

Despite many advantages, CDA is still in its infancy. One notable limitation of CDA is that it requires a sufficient number of APs as reference points and their perfect locations. There are several promising directions to overcome this limitation, all of which could be interesting topics for future research. First, following the principles in multi-point positioning \cite{Zhang2021}, it is possible to jointly utilize heterogeneous APs in different communication protocols, such as WiFi, UWB, and Bluetooth, to secure a sufficient number of APs. Second, another promising way is to use imperfect AP locations while running distributed learning \cite{Park2021} over the DP part of CDA, which can filter out the noise in AP locations. Last but not least, leveraging dominant reflectors and their associated PEL clusters as landmarks makes it viable to design CDA-based simultaneous localization and mapping \cite{Ge2020}, through which AP locations can be identified in real time. 

\section*{Appendix}

\subsection{Verification of RTT-Sum Hypothesis}

\begin{figure}[t]
\centering
{
\includegraphics[width=8.50cm]{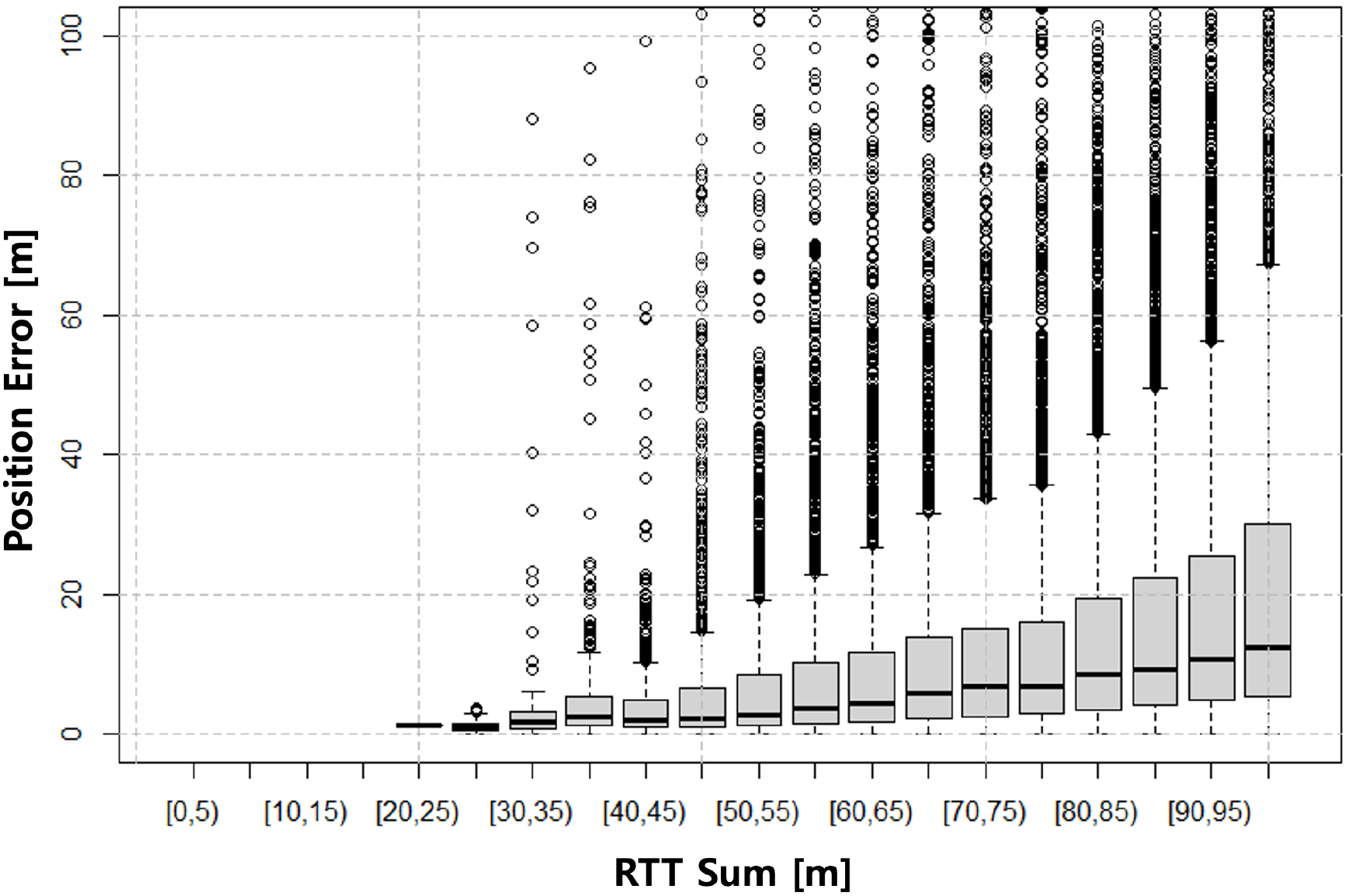}
}
\caption{The relation between RS $\beta(\ell)$ of \eqref{RS_Def} and position error represented by a boxplot. Each box's lower, middle, and upper horizontal lines represent the first, median, and third quartiles. The whisker above the box represents the maximum value of the data samples. Circles represent the outliers located outside the allowable range.} \label{HypothesisVerify}
\end{figure}

In this subsection, we aim to verify the hypothesis that PELs with smaller RTT sum $\beta(\ell)$ are located much closer to the ground truth. As shown in Fig. \ref{HypothesisVerify}, it is observed that smaller $\beta(\ell)$ is likely to provide a PEL with both median and 
variance being smaller. On the other hand, the number of outliers represented by circles increases as $\beta(\ell)$ becomes higher. It is thus concluded that the hypothesis makes sense. 

\subsection{Effect of Parameter Configuration}
This subsection attempts to explain the effect of two key parameters, say $M$ and $q$, representing the number of APs required for one PEL and the portion of the remaining PELs after tandem filtering. As mentioned in Remark \ref{Remark:ParamConfig}, the relation between $M$ and $q$ is given as $q=M^\delta$, where the probability of a clean RTT $\delta$ is approximately $0.465$ at the concerned experiment site. Given $M$, the parameter $q$ is accordingly determined. Table \ref{Table:ParamConfig} summarizes the average and standard deviations of the location estimate errors when using different pairs of $M$ and $q$, verifying the optimality of the current configuration of $(M,q)=(3,0.1)$.

\begin{table}[]
\centering
\caption{Performance Comparison among Different Parameter Configurations of $M$ and $q$}
\begin{tabular}{@{}l|c|c|c|c@{}}
\toprule
$(M,q)$ & \multicolumn{1}{l|}{$(3,0.1)$} & \multicolumn{1}{l|}{$(4,0.05)$} & \multicolumn{1}{l|}{$(5, 0.02)$} & \multicolumn{1}{l}{$(6,0.01)$} \\ \midrule
$\mathsf{avg}$ (m) & {$1.602$}                        & $1.886$                         & $2.064$                        & $2.258$                         \\ \midrule
$\mathsf{std}$  (m) &  {$1.156$}                        & $1.497$                         & $1.754$                        & $2.002$                         \\ \bottomrule
\end{tabular}
\label{Table:ParamConfig}
\end{table}

\subsection{Derivation of Kalman Gain}

Given the ME $\boldsymbol{y}_{\textrm{RE \& RS}}$ specified in \eqref{LocEst_Filtering} with its covariance matrix $\boldsymbol{R}\in\mathbb{R}^{2 \times 2}$ and the initial position estimate $\boldsymbol{y}_0$, 
we set an initial state vector as $\boldsymbol{p}_0=[\boldsymbol{y}_0;1]^T\in \mathbb{R}^{3 \times 1}$, where the last element~$1$ is a dummy variable for the following linear representation; The PE state can be expressed as 
$\boldsymbol{p}_{\textrm{PE}}= \boldsymbol{A}\boldsymbol{p}_0$ with the matrix $\boldsymbol{A}=
\begin{bmatrix}
    1 & 0 & v_x  \\
    0 & 1 & v_y  \\
    0 & 0 &1 \\
\end{bmatrix},$
where $v_x$ and $v_y$ are projected movement onto $x$ and $y$ axes, specified in Sec. \ref{subsection:Cov_Update}. The  covariance matrix of $\boldsymbol{p}_{\textrm{PE}}$ is given as $\boldsymbol{Q}$.
Next, following the theory of KF, the final state, denoted by $\boldsymbol{p}_{\textrm{KF}}=[\boldsymbol{y}_{\textrm{KF}};1]^T$,  is given as
\begin{align}
\boldsymbol{p}_{\textrm{KF}}&=\boldsymbol{p}_{\textrm{PE}}+\boldsymbol{B}(\boldsymbol{y}_{\textrm{RE \& RS}}-\boldsymbol{H}\boldsymbol{p}_{\textrm{PE}})\nonumber\\
&=\boldsymbol{B}\boldsymbol{y}_{\textrm{RE \& RS}}+(\boldsymbol{I}_3-\boldsymbol{B}\boldsymbol{H})\boldsymbol{p}_{\textrm{PE}},
\end{align}
where $\boldsymbol{B}\in\boldsymbol{R}^{3 \times 2}$ is the Kalman gain based on the above $3$-by-$1$ state representation and the matrix $\boldsymbol{H}=\begin{bmatrix}
    1 & 0 & 0  \\
    0 & 1 & 0  \\
\end{bmatrix}$ is used for integrating two position estimates with different sizes. Last, we can reduce the above equation into \eqref{KF} discarding the third element, where $\boldsymbol{G}$ in \eqref{KF} is a $2$-by-$2$ matrix whose first and second row vectors are those of $\boldsymbol{B}$. 

\subsection{Another Field Experiment}\label{AdditionalExperiment}

\begin{figure}[t]
\centering
{
\includegraphics[width=8.90cm]{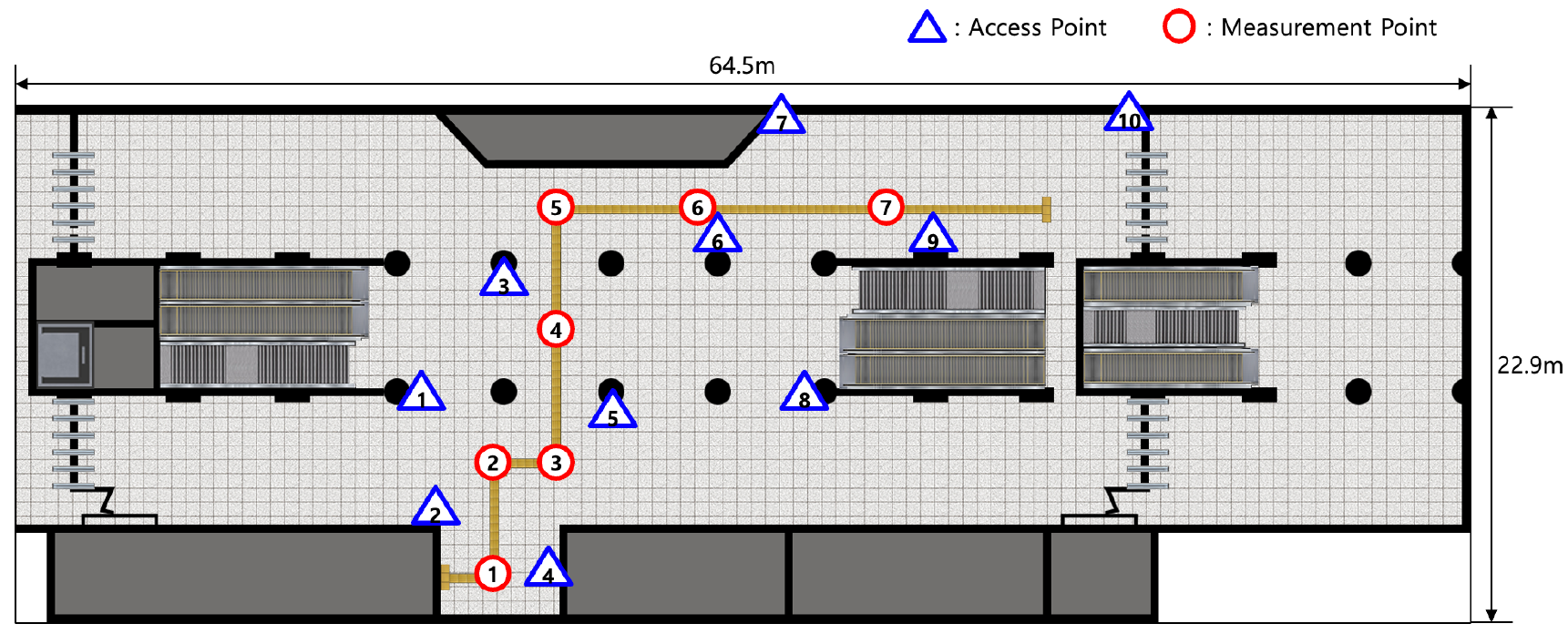}
}
\caption{Floor plan of the experiment site (Geumnamno4-ga subway station B3F, Gwangju, South Korea). The detailed experiment settings are described in Appendix D.} \label{Floor_Plan_2}
\end{figure}

We conducted another field experiment at Geumnamno4-ga subway station B3F, Gwangju, South Korea. The experiment settings are equivalent to the experiment in Sec. \ref{Sec3} unless specified. Each experiment comprises $7$ MPs. We repeated this experiment $9$ times for one hour. {The experimental site's floor plan and results are given in Fig. \ref{Floor_Plan_2} and Table \ref{table:RERS_additional}, respectively.}  

\begin{table*}\centering
\caption{Results of the experiments in Appendix D. }\label{table:RERS_additional}
\begin{tabular}{ll|ccccccccc|c}
\toprule
\multicolumn{2}{c|}{{Experiment \#}}                                                                                                         & $1$             & $2$    & $3$    & $4$    & $5$    & $6$    & $7$    & $8$    & $9$    & {Total} \\ \midrule
\multicolumn{1}{c|}{\multirow{2}{*}{{LLS-RS \cite{Guvenc2008}}}}                                                                                  & $\mathsf{avg}$ (m) & $4.51$ & $4.27$ & $4.84$ & $4.19$ & $4.61$ & $5.04$ & $5.08$ & $4.09$ & $3.43$ & $4.45$           \\ 
\multicolumn{1}{c|}{}                                                                                                                  & $\mathsf{std}$ (m) & $1.70$          & $1.60$ & $2.89$ & $1.84$ & $2.40$ & $2.76$ & $2.82$ & $1.42$ & $1.33$ & $2.08$            \\ \midrule
\multicolumn{1}{c|}{\multirow{2}{*}{{CDA only}}}                                                                              & $\mathsf{avg}$ (m) & $1.82$          & $1.67$ & $1.80$ & $1.91$ & $2.10$ & $1.64$ & $1.43$ & $1.33$ & $1.64$ & $1.70$            \\ 
\multicolumn{1}{c|}{}                                                                                                                  & $\mathsf{std}$ (m) & $1.01$          & $0.83$ & $1.53$ & $1.05$ & $1.99$ & $1.68$ & $0.85$ & $1.02$ & $0.84$ & $1.20$            \\ \midrule
\multicolumn{1}{c|}{\multirow{2}{*}{{\begin{tabular}[c]{@{}c@{}}CDA \& PDR\\ (KF w/ deterministic $\boldsymbol{R}$ \& $\boldsymbol{Q}$)\end{tabular}}}} & $\mathsf{avg}$ (m) & $1.76$          & $1.24$ & $1.42$ & $1.86$ & $1.85$ & $1.15$ & $0.97$ & $0.95$ & $1.30$ & $1.39$            \\ 
\multicolumn{1}{c|}{}                                                                                                                  & $\mathsf{std}$ (m) & $0.71$          & $0.71$ & $1.26$ & $0.78$ & $1.47$ & $0.94$ & $0.71$ & $0.59$ & $0.47$ & $0.85$            \\ \midrule
\multicolumn{1}{c|}{\multirow{2}{*}{{\begin{tabular}[c]{@{}c@{}}CDA \& PDR\\ (KF w/ updating $\boldsymbol{R}$ \eqref{R_Update} \& $\boldsymbol{Q}$ \eqref{Q_Update})\end{tabular}}}}     & $\mathsf{avg}$ (m) & $1.71$          & $1.26$ & $1.31$ & $1.95$ & $0.88$ & $0.98$ & $0.86$ & $1.60$ & $1.52$ & $1.34$            \\ 
\multicolumn{1}{c|}{}                                                                                                                  & $\mathsf{std}$ (m) & $0.94$          & $0.69$ & $0.50$ & $0.78$ & $0.49$ & $0.54$ & $0.60$ & $0.98$ & $0.63$ & $0.68$            \\ \bottomrule
\end{tabular}\end{table*}

\bibliographystyle{ieeetr}
\bibliography{CDA_Mag}

\begin{thebibliography}{10}

\bibitem{CDA2022_VTC}
S.~M. Yu, J.~Park, and S.-W. Ko, ``Combinatorial data augmentation for
  real-time indoor positioning: Concepts and experiments,'' in {\em Proc. IEEE
  VTC 2022 Spring}, (Helsinki, Finland), pp.~1--5, Jun. 2022.

\bibitem{Bensky2016}
A.~Bensky, {\em Wireless positioning technologies and applications}.
\newblock Artech House, 2016.

\bibitem{Laoudias2018}
C.~{Laoudias}, A.~{Moreira}, S.~{Kim}, S.~{Lee}, L.~{Wirola}, and
  C.~{Fischione}, ``A survey of enabling technologies for network localization,
  tracking, and navigation,'' {\em IEEE Commun. Surveys Tuts.}, vol.~20,
  pp.~3607--3644, fourthquarter 2018.

\bibitem{Zafari2019}
F.~Zafari, A.~Gkelias, and K.~K. Leung, ``A survey of indoor localization
  systems and technologies,'' {\em IEEE Commun. Surveys Tuts.}, vol.~21,
  pp.~2568--2599, thirdquarter 2019.

\bibitem{Fischer2014}
S.~Fischer, ``Observed time difference of arrival ({OTDOA}) positioning in
  {3GPP} {LTE},'' {\em Qualcomn White Paper}, Jun. 2014.

\bibitem{Han2021}
K.~Han, S.~M. Yu, S.-L. Kim, and S.-W. Ko, ``Exploiting user mobility for
  {WiFi} {RTT} positioning: A geometric approach,'' {\em IEEE Internet Things
  J.}, vol.~8, pp.~14589--14606, Apr. 2021.

\bibitem{Feng2020}
D.~Feng, C.~Wang, C.~He, Y.~Zhuang, and X.-G. Xia, ``Kalman-filter-based
  integration of {IMU} and {UWB} for high-accuracy indoor positioning and
  navigation,'' {\em IEEE Internet Things J.}, vol.~7, pp.~3133--3146, Apr.
  2020.

\bibitem{Banin2017}
L.~Banin, O.~Bar-Shalom, N.~Dvorecki, and Y.~Amizur, ``High-accuracy indoor
  geolocation using collaborative time of arrival-whitepaper,'' {\em IEEE
  802.11-17/1397R0}, Sep. 2017.

\bibitem{Ibrahim2018}
M.~Ibrahim, H.~Liu, M.~Jawahar, V.~Nguyen, M.~Gruteser, R.~Howard, B.~Yu, and
  F.~Bai, ``Verification: Accuracy evaluation of {WiFi} fine time measurements
  on an open platform,'' in {\em Proc. 24th Annu. Int. Conf. Mobile Comput.
  Netw. (MobiCom)}, pp.~417--427, 2018.

\bibitem{miao2007positioning}
H.~Miao, K.~Yu, and M.~Juntti, ``Positioning for {NLOS} propagation: Algorithm
  derivations and {C}ramer-{R}ao bounds,'' {\em IEEE Trans. Veh. Tech.},
  vol.~56, pp.~2568--2580, Sep. 2007.

\bibitem{han2018sensing}
K.~Han, S.-W. Ko, H.~Chae, B.-H. Kim, and K.~Huang, ``Hidden vehicle sensing
  via asynchronous {V2V} transmission: {A} multi-path-geometry approach,'' {\em
  IEEE Access}, vol.~7, pp.~pp.169399--169419, Dec. 2019.

\bibitem{Ko2021_VP_Mag}
S.-W. Ko, H.~Chae, K.~Han, S.~Lee, D.-W. Seo, and K.~Huang, ``{V2X}-based
  vehicular positioning: Opportunities, challenges, and future directions,''
  {\em IEEE Wireless Comm.}, vol.~28, pp.~144--151, Apr. 2021.

\bibitem{Yang2015}
Z.~Yang, C.~Wu, Z.~Zhou, X.~Wang, and Y.~Liu, ``Mobility increases
  localizability: A survey on wireless indoor localization using inertial
  sensors,'' {\em ACM Comput. Surveys}, vol.~47, pp.~54:1--54:34, Apr. 2015.

\bibitem{Zhang2020}
L.~Zhang, D.~Sidoti, A.~Bienkowski, K.~R. Pattipati, Y.~Bar-Shalom, and D.~L.
  Kleinman, ``On the identification of noise covariance and adaptive {Kalman}
  filtering: A new look at a 50 year-old problem,'' {\em IEEE Access}, vol.~8,
  pp.~59362 -- 59388, Mar. 2020.

\bibitem{Marano2010}
S.~Marano, W.~M. Gifford, H.~Wymeersch, and M.~Z. Win, ``{NLOS} identification
  and mitigation for localization based on {UWB} experimental data,'' {\em IEEE
  J. Sel. Areas Commun.}, vol.~28, pp.~1026--1035, Sep. 2010.

\bibitem{Choi2018}
J.-S. Choi, W.-H. Le, J.-H. Lee, J.-H. Lee, and S.-C. Kim, ``Deep learning
  based {NLOS} identification with commodity {WLAN} devices,'' {\em IEEE Trans.
  Veh. Techno.}, vol.~67, pp.~3295--3303, Apr. 2018.

\bibitem{Huang2020}
C.~Huang, A.~F. Molisch, R.~He, R.~Wang, P.~Tang, B.~Ai, and Z.~Zhong,
  ``Machine learning-enabled {LOS/NLOS} identification for {MIMO} systems in
  dynamic environments,'' {\em IEEE Trans. Wireless Commun.}, vol.~19,
  pp.~3643--3657, Jun. 2020.

\bibitem{Vo2015}
Q.~D. Vo and P.~De, ``A survey of fingerprint-based outdoor localization,''
  {\em IEEE Commun. Surveys Tuts.}, vol.~18, no.~1, pp.~pp. 491--506, 2015.

\bibitem{Yiu2017}
S.~Yiu, M.~Dashti, H.~Claussen, and F.~P. Cruz, ``Wireless {RSSI}
  fingerprinting localization,'' {\em Elsevier Signal Processing}, vol.~131,
  pp.~235--244, Feb. 2017.

\bibitem{Wang2016TVT}
X.~Wang, L.~Gao, S.~Mao, and S.~Pandey, ``{CSI}-based fingerprinting for indoor
  localization: A deep learning approach,'' {\em IEEE Trans. Veh. Techno.},
  vol.~66, pp.~763--776, Mar. 2017.

\bibitem{Hashem2020}
O.~Hashem, M.~Youssef, and K.~A. Harras, ``{WiNar}: {RTT}-based sub-meter
  indoor localization using commercial devices,'' in {\em Proc. IEEE Intern.
  Conf. Pervasive Comput. Commun.}, 2020.

\bibitem{Zhou2021}
C.~Zhou, J.~Liu, M.~Sheng, Y.~Zheng, and J.~Li, ``Exploiting fingerprint
  correlation for fingerprint-based indoor localization: A deep learning based
  approach,'' {\em IEEE Trans. Veh. Techno.}, vol.~70, pp.~5762--5774, Jun.
  2021.

\bibitem{Li2019}
L.~Li, X.~Guo, N.~Ansari, and H.~Li, ``A hybrid fingerprint quality evaluation
  model for wifi localization,'' {\em IEEE Internet Things J.}, vol.~6,
  pp.~9829--9840, Dec. 2019.

\bibitem{Chen2016}
L.~Chen, K.~Yang, and X.~Wang, ``Robust cooperative {Wi-Fi} fingerprint-based
  indoor localization,'' {\em IEEE Internet Things J.}, vol.~3, pp.~1406--1417,
  Dec. 2016.

\bibitem{Wang2016}
B.~Wang, Q.~Chen, L.~T. Yang, and H.-C. Chao, ``Indoor smartphone localization
  via fingerprint crowdsourcing: Challenges and approaches,'' {\em IEEE
  Wireless Commun.}, vol.~23, pp.~82--89, Jun. 2016.

\bibitem{Caso2015}
G.~Caso and L.~De~Nardis, ``On the applicability of multi-wall multi-floor
  propagation models to wifi fingerprinting indoor positioning,'' in {\em in
  Proc. Int. Conf. Future Access Enablers for Ubiquitous and Intell.
  Infrastructures}, pp.~166--172, Springer, 2015.

\bibitem{Renaudin2018}
O.~Renaudin, T.~Zemen, and T.~Burgess, ``Ray-tracing based fingerprinting for
  indoor localization,'' in {\em in Proc. IEEE SPAWC}, 2018.

\bibitem{Silva2022}
G.~M. Mendoza-Silva, A.~C. Costa, J.~Torres-Sospedra, M.~Painho, and J.~Huerta,
  ``Environment-aware regression for indoor localization based on wifi
  fingerprinting,'' {\em IEEE Sensors J.}, vol.~22, pp.~4978--4988, Mar. 2022.

\bibitem{Guvenc2008}
I.~Guvenc, S.~Gezici, F.~Watanabe, and H.~Inamura, ``Enhancements to linear
  least squares localization through reference selection and {ML} estimation,''
  in {\em IEEE WCNC}, 2008.

\bibitem{Qiao2014}
T.~Qiao and G.~Liu, ``Improved least median of squares localization for
  non-line-of-sight mitigation,'' {\em IEEE Commun. Lett.}, vol.~18,
  pp.~1451--1454, Aug. 2014.

\bibitem{Chen1999}
P.-C. Chen, ``A non-line-of-sight error mitigation algorithm in location
  estimation,'' in {\em Proc. IEEE WCNC}, 1999.

\bibitem{Olone2018}
C.~E. O'Lone, H.~S. Dhillon, and R.~M. Buehrer, ``A statistical
  characterization of localization performance in wireless networks,'' {\em
  IEEE Trans. Wireless Commun.}, vol.~17, pp.~5841--5856, Sep. 2018.

\bibitem{Zhang2021}
Z.~Zhang, S.-W. Ko, R.~Wang, and K.~Huang, ``Cooperative multi-point vehicular
  positioning using millimeter-wave surface reflection,'' {\em IEEE Trans.
  Wireless Commun.}, vol.~20, pp.~2221--2236, 2021.

\bibitem{Park2021}
J.~Park, S.~Samarakoon, A.~Elgabli, J.~Kim, M.~Bennis, S.-L. Kim, and
  M.~Debbah, ``Communication-efficient and distributed learning over wireless
  networks: Principles and applications,'' {\em Proc. of IEEE}, vol.~109,
  pp.~796--819, May 2021.

\bibitem{Ge2020}
Y.~Ge, H.~Kim, F.~Wen, S.~Kim, and H.~Wymeersch, ``Exploiting diffuse multipath
  in {5G} {SLAM},'' in {\em Proc. GLOBECOM}, 2020.

\end{thebibliography}

\end{document}